\documentclass {aa}
\usepackage{graphicx}
\usepackage{txfonts}
\begin{document}
   \title{Searching for Cool Core Clusters at High redshift}

   \author{Joana S. Santos,
          \inst{1}
          Piero Rosati,\inst{2}
          Paolo Tozzi,\inst{3}
          Hans B\"ohringer,\inst{1}
          Stefano Ettori \inst{4}
	  and Andrea Bignamini\inst{3}
	}

   \offprints{J. Santos, jsantos@mpe.mpg.de}

   \institute{
	     \inst{1} Max-Planck-Institut f\"ur extraterrestrische Physik,
              Giessenbachstra\ss e, 85748 Garching, Germany\\
             \email{jsantos@mpe.mpg.de} \\
             \inst{2} European Southern Observatory, Karl-Schwarzchild Strasse 2, 85748 
Garching, Germany \\
	     \inst{3} INAF, Osservatorio Astronomico di Trieste, via G.B. Tiepolo 11, 
34131, Trieste,Italy \\
             \inst{4} INAF, Osservatorio Astronomico di Bologna, via Ranzani 1, 
40127, Bologna, Italy
             }

   \date{Received ... ; accepted ...}

 
  \abstract
   {}
   {We investigate the detection of Cool Cores (CCs) in the distant galaxy cluster population,
   with the purpose of measuring the CC fraction out to redshift $0.7\le z<1.4$. Using a sample 
   of nearby clusters spanning a wide range of morphologies, we define criteria to characterize 
   cool cores, which are applicable to the high redshift sample. }
   {We analyzed azimuthally averaged surface brightness (SB) profiles using the known scaling 
   relations and fitted single/double $\beta$ models to the data. Additionally, we measured a 
   surface brightness concentration, $c_{SB}$, as the ratio of the peak over the ambient 
   SB. To verify that this is an unbiased parameter as a function of redshift, we developed a 
   model independent "cloning" technique to simulate the nearby clusters as they would appear 
   at the same redshifts and luminosities as those in the distant sample. This method is based 
   on the application of the cosmological surface brightness dimming to high resolution Chandra 
   images and assumes no intrinsic cluster evolution. 
    A more physical parameterization of CC presence is obtained by computing the cooling time at 
a radius of 20 kpc from the cluster center.}
   {The distribution of the SB concentration and the stacked radial profiles of the low-$z$ sample, 
   combined with published information on the CC properties of these clusters, show 3 degrees 
   of SB cuspiness: non-CC, moderate and strong CC. The same analysis applied to the high-$z$ clusters
   reveals two regimes: non-CC and moderate CC. The cooling time distribution corroborates this result
   by showing a strong negative correlation with $c_{SB}$. }
   {Our analysis indicates a significant fraction of distant clusters 
   harboring a moderate CC out to $z$=1.4, similar to those found in the local sample. The absence of strong 
   cooling which we report is likely linked with a higher merger rate expected at redshift $z > 0.7$,  
   and should also be related with the shorter age of distant clusters, implying less time to develop 
   a cool core.}

   \keywords{Galaxy clusters - cosmology: Galaxy clusters - high redshift: observations - X-rays 
               }

   \authorrunning{J.S.Santos et al.}
   \titlerunning{Searching for Cool Core Clusters at High redshift}

   \maketitle
%

\section{Introduction}

Galaxy clusters are filled with hot, diffuse gas - the intra cluster medium (ICM) - which 
strongly emits in X-rays. While the cluster dark matter distribution is generally described 
by its gravitational potential (Navarro, Frenk \& White 1997), the physical processes which 
govern the behavior of the ICM are complex and include non-gravitational phenomena. Cool cores
 are a manifestation of these intricate processes.

According to observations (Jones \& Forman \cite{jones}, Fabian et al. \cite{fabian94}, Peres et 
al. \cite{peres}, Chen et al. \cite{chen}), relaxed galaxy clusters are most often found to exhibit 
at their centers a significant drop in the ICM gas temperature due to radiative cooling. It is by now 
established that the cooling time in the central regions of clusters is shorter than the
Hubble time, thus originating very dense cores. Therefore, we expect to observe a surface 
brightness (SB) peak in these regions (with a typical radius of the order of 70 kpc - see 
for example Vikhlinin et al. \cite{vikhlinin05}) along with some other features such as an 
enhanced iron abundance. 

The well known cooling flow problem stems from the observation that the few detected emission
 lines are not as strong as expected to justify the predicted cooling rate. Further 
observational evidence is found in the ratio of the central temperature with respect to the 
global cluster temperature $(T_{central}/T)$, which remains at a factor $\thicksim$1/3 
(Peterson et al. \cite{peterson03}, Bauer et al. \cite{bauer}), and the mass deposition rates 
(\.{M}) are much smaller than expected (Edge \& Frayer 2003).

The existence of a feedback mechanism which counteracts cooling is now widely accepted and 
currently, the most debated picture is heating by active galactic nuclei (AGN) (see review by 
Fabian et al. \cite{fabian84}). Although several plausible models exist 
which may explain how this mechanism works (Churazov et al. \cite{churazov}, Br\"uggen \& Kaiser 2002),  
and there is evidence for enough output mechanical energy to suppress cooling, it remains 
unclear how this energy is distributed in a homogenous way, such that cool cores appear in the
form observed. Nevertheless, the CC - AGN connection has been unambiguously demonstrated with 
observations, e.g., Burns \cite{burns90}, Eilek \cite{eilek}, Sanderson et al. \cite{sanderson} 
report that nearly every cool core cluster hosts a radio emitting AGN, creating cavities in 
the X-ray gas. 

There is clear evidence for clusters harboring cool cores, however, many clusters do not
 show any signatures of cooling. At this point, it may be pertinent to question 
whether there is a CC/Non CC bimodal cluster population. In this the case, are CC clusters
primordial and were some of them (the non-cool core clusters) have been wiped out by mergers
 or an AGN heating overshoot? Conversely, some authors support a scenario in which non-CC 
evolve to CC if no major mergers occur (O'Hara et al. \cite{o'hara}).

The role of mergers in destroying cool cores has long been debated (Fabian et al. \cite{fabian84}),
 both from observational results and predictions from simulations. Currently, observations 
seem to favor CC destruction through cluster mergers (Allen et al. \cite{Allen}, Sanderson et al. \cite{sanderson}). Simulations, however, yield ambivalent results. Using adaptive mesh
 refinement simulations, Burns et al. \cite{burns} advocate that non-cool core clusters undergo major 
mergers which destroy embryonic cool cores while CC clusters grow at a slower rate without 
suffering early substantial mergers. However, an analysis of smoothed particle hydrodynamics 
simulations of X-ray cluster mergers by Poole et al. \cite{poole}, suggests that the heating of the 
ICM arising from mergers is not sufficient to prevent cooling. They argue that, in the 
$\Lambda CDM$ scenario, the merger rate is too small to account for the local abundance of 
non-CC clusters whose disruptions occurs on short timescales. We note, however, that 
a good agreement in the overall properties of observed and simulated CCs/NCCs is found by Burns et al., 
who were able to qualitatively reproduce the temperature structure observed in cool cores.

    The present observational census on the abundance of cool core clusters sets a fraction 
of Cool Core Clusters (CCC) in the local Universe in the range 50 -70 \% (Peres et al. \cite{peres}, 
Chen et al. \cite{chen}). As we move to increasing redshifts, results on this topic become more sparse.
     Using spatially resolved spectroscopy, Bauer et al. \cite{bauer} presented a comparative study 
of cool cores at low and intermediate redshift (mean $z$=0.06 and 0.22, respectively). The cool 
cores at intermediate redshift show the same temperature decrement, $T/T_{central}\thicksim$ 3-4, 
as the nearby CC's, and have a frequency rate similar to the local one. The central cooling time 
measurements confirmed this result. Consequently, they find no signs of evolution of the cool 
core fraction in the redshift range $z\thicksim~0.15-0.4$ to now.

     At present, the knowledge of the high redshift cluster population is still rather poor: 
only 7 X-ray selected clusters with $z>1$ have been confirmed (Rosati et al. \cite{rosati99}, Perlman et 
al. \cite{perlman}, Stanford et al. \cite{stanford02}, Mullis et al. \cite{mullis}, Stanford et al. \cite{stanford06}). 
In particular, determining and quantifying the existence of cool cores at high redshift 
remains largely unexplored due to the lack of photon collecting power of current instruments to 
provide spectra with sufficient statistics for distant CC's. A first attempt to characterize a 
peaked core in high redshift systems was undertaken in Ettori et al. \cite{ettori}, using double beta 
model fitting. The outcome of this analysis states that a second beta model component cannot 
be required to describe the distant cluster data - at least, not in the same way as observed for
 nearby cool core clusters. Recently, Vikhlinin et al. \cite{vikhlinin06} presented a study on the evolution of
 CCC at $z>0.5$ based on pure imaging data, using a distant cluster sample drawn from the 400SD
 survey imaged with Chandra. These authors claim a strong evolution of the CC fraction in this redshift
 range, implying that we are missing cooling flows in the distant cluster population. 
However, this claim has been recently challenged with simulations by Burns et al. \cite{burns}, 
who found no evidence for such a decline in the Cool Core fraction in simulated data with 
redshift up to one.

    The understanding of the formation and evolution of cooling flows is fundamental in 
pursuing cosmological studies using galaxy clusters. It is already known that cool core 
clusters affect the scaling relations causing a departure from the theoretical expectations 
and increasing the scatter (Fabian et al. \cite{fabian94}, Zhang et al. \cite{zhang}). Whether or not we can put
 together a consistent picture of clusters with and without cool cores will greatly affect the
 reliability of cluster results in deriving cosmological parameters.

    In this paper we investigate the fraction of cool core clusters in a high 
redshift sample imaged with Chandra at $0.7\le z<1.4$. To this aim we define and quantify 
criteria to select clusters as CCC in a sample of nearby clusters, $0.15<z<0.3$, which 
are then applied to the distant sample. The low-$z$ clusters exhibit markedly different 
morphologies which is essential to characterize cool cores with varying strength. The low 
photon statistics of the distant clusters made a spatially resolved temperature analysis 
inviable, constraining our study to a spatial analysis of the cluster images. We analyzed 
the overall surface brightness and determined cooling times in the clusters' central regions. 
The surface brightness diagnostics were developed based on the nearby sample and were 
validated to high-$z$ using simulated distant clusters. These were obtained with a novel simulation 
method which we present here, the cloning technique, in which we simulate distant galaxy clusters 
by emulating the appearance of the low-$z$ clusters at the same redshifts of the observed distant 
sample. Such simulations allow us to make an unbiased comparison of cluster spatial properties 
between nearby and distant populations.

   The two cluster samples are drawn from the Chandra archive. The high-$z$ sample, which is 
the most relevant for this study, is a statistically complete sample based on the Rosat 
Distant Cluster Sample (RDCS), from Rosati et al. \cite{rosati98}. This dataset allows us to draw conclusions
 on the abundance of CCC when the Universe was $\thicksim 1/3$ of its age. On the other hand, 
the nearby sample was constructed by taking morphologically interesting archive clusters and was 
used as a test set to understand how we can characterize cool cores. 

We note that archive samples will likely be biased in the sense that the majority
of the observed clusters are "interesting", where interesting here means showing signs of 
mergers / disruptions, or on the contrary, having very regular shapes. In addition, cluster 
selection based on flux limited samples will preferentially find high luminosity/mass 
clusters at high redshift. This effect, goes in the direction of favoring CC clusters (which 
have a bright core) with increasing $z$. For these reasons, constructing complete, unbiased 
samples to address issues of cluster evolution presents difficulties which should be carefully 
dealt with. Since the distant and nearby samples are not drawn from a homogenous sample with 
a known selection function, we do not address the issue of evolution of cool cores on this paper.

     As a complement to this X-ray study and with the aim of understanding sample 
selection effects, we shortly investigate the surface brightness properties of an 
optically selected sample of high-$z$ galaxy clusters imaged with Chandra. 
We assess the incidence rate of cool cores and compare it with the results on the X-ray 
selected cluster set. 

     The paper is organized in the following manner: the cluster samples are introduced in 
section 2 and in section 3 the surface brightness profiles are presented, together with the 
beta model fitting. In section 4 we introduce the surface brightness concentration parameter 
and present the simulation method we used to test its application. In section 5 we present the cooling time analysis.
Section 7 is dedicated to a comparison of X-ray surface brightness properties in 
optically and X-ray selected high redshift galaxy clusters. Finally, in section 8 we discuss the 
results obtained and conclude.

     The cosmological parameters used throughout the paper are: $H_{0}$=70 km/s/Mpc, 
$\Omega_{\Lambda}$=0.7 and $\Omega_{\rm m}$ =0.3.

\section{Cluster samples}

\begin{figure*} 
\centering
\includegraphics[height=7cm,angle=0,clip=true]{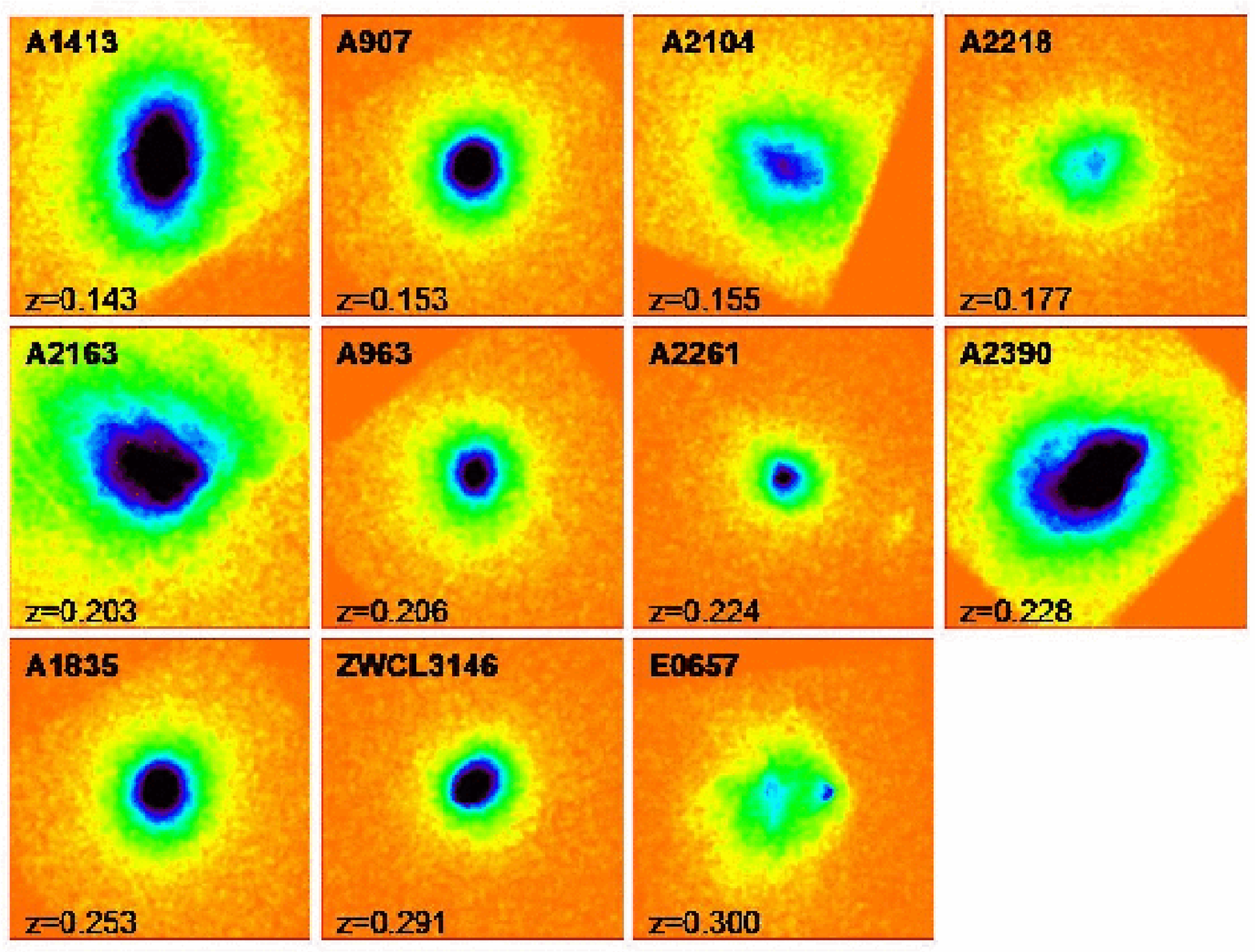}
\caption{Nearby cluster sample observed with Chandra (point sources are removed). Images are 
smoothed with a gaussian of sigma=3 and rescaled by the square root of the intensity. Individual 
boxes have a size of 8x8 arcmin. Cluster redshift increases from top-left to bottom-right.}
 \label{Figlowzsample}
\end{figure*}

\begin{figure*} 
\centering
\includegraphics[width=11cm,height=10cm,angle=0,clip=true]{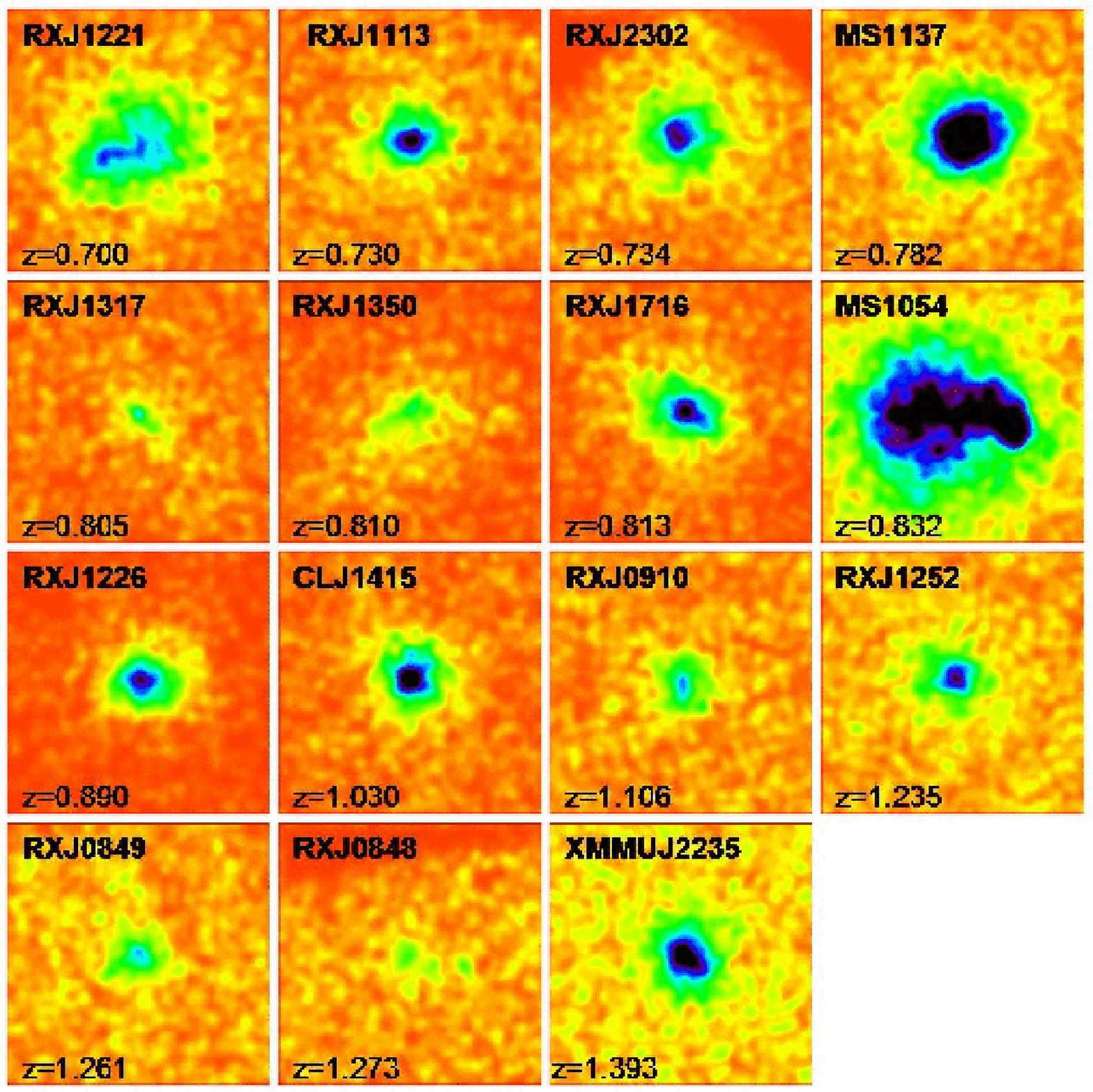}
\caption{X-ray selected, distant cluster sample observed with Chandra (point sources are removed). Images are 
smoothed with a gaussian of sigma=3 and rescaled by the square root of the intensity. Individual 
boxes have a size of 3x3 arcmin. Cluster redshift increases from top-left to bottom-right.}
 \label{Fighighzsample}
\end{figure*}

 We used archive Chandra ACIS-I and ACIS-S images in the [0.5-5.0] keV energy range: 
the 11 low-$z$ clusters (Table 1) were selected from the sample compiled by Baldi et al. \cite{baldi}, 
and the high redshift X-ray  selected sample, which comprises 15 clusters (first part of 
Table 2), was drawn from the sample presented in Balestra et al. \cite{balestra}. We therefore 
refer the reader to these two papers for details in the data processing. In Table 2 we also 
present the optically selected distant sample which will be discussed in section 7.

\begin{table*}
\caption{Low redshift cluster sample: cluster name (1), redshift (2), spectral temperature (3),
 Right Ascension (4), Declination (5) and net exposure time (6)}             
\label{table:1}      
\centering                          
\begin{tabular}{c c c c c c}        
\hline\hline                 
Cluster ID$^{(1)}$ & z$^{(2)}$ & T (keV)$^{(3)}$ & RA$^{(4)}$ & DEC$^{(5)}$  & Exposure (ksec)$^{(6)}$ \\  
\hline                        
A1413$^{\mathrm a}$ 	 & 0.143 & $7.52^{+0.20}_{-0.12}$ & 11:55:18.20	& +23:24:28.80  & 136 \\
A907$^{\mathrm a}$	 & 0.153 & 5.82 $\pm$0.12	  & 09:58:22.00	& -11:03:50.40  & 100 \\ 
A2104$^{\mathrm a}$	 & 0.155 & 6.76	$\pm$0.19         & 15:40:07.90	& -03:18:04.00  &  48 \\
A2218$^{\mathrm a}$	 & 0.177 & 6.25 $\pm$0.31	  & 16:35:56.00	& +66:12:45.00  &  57 \\
A2163$^{\mathrm b}$	 & 0.203 & $12.3^{+1.3}_{-1.1}$	  & 16:15:45.77	& -06:08:55.00  &  80 \\
A963$^{\mathrm a}$ & 0.206 & $6.02^{+0.28}_{-0.19}$	  & 10:17:03.40	& +39:02:51.00  &  36 \\
A2261$^{\mathrm a}$	 & 0.224 & $7.43^{+0.49}_{-0.27}$ & 17:22:27.20	& +32:07:58.00  &  32 \\
A2390$^{\mathrm a}$	 & 0.228 & 9.35 $\pm$0.15 	  & 21:53:36.50	& +17:41:45.00  & 111 \\
A1835$^{\mathrm a}$	 & 0.253 & 8.09 $\pm$0.53	  & 14:01:02.00	& +02:52:40.00  &  30 \\
ZwCl3146$^{\mathrm a}$ & 0.291 & 8.59 $\pm$0.39	 	  & 10:23:39.60	& +04:11:10.00  &  45 \\
E0657$^{\mathrm c}$ & 0.300 & 14.1	$\pm$0.2          &  6:58:28.60	& -55:56:36.03  &  25 \\
\hline                                  
\end{tabular}
\begin{list}{}{}
\item$^{\mathrm[a]}$ Temperatures taken from Baldi et al. \cite{baldi}, $^{\mathrm[b]}$ temperature 
from Markevitch et al. \cite{markevitch01}, $^{\mathrm[c]}$ temperature from Markevitch et al. \cite{markevitch05}.
\end{list}
\end{table*}

\begin{table*}
\caption{High redshift cluster samples: cluster name (1), redshift (2), spectral temperature (3),
 Right Ascension (4), Declination (5) and net exposure time (6)}             
\label{table:2}      
\centering           
\begin{tabular}{c c c c c c} 
\hline\hline                
Cluster ID$^{(1)}$ & z$^{(2)}$ & T (keV)$^{(3)}$ & RA$^{(4)}$ & DEC$^{(5)}$  & Exposure (ksec)$^{(6)}$ \\   
\hline                        

RX J1221+4918	  & 0.700     & $8.4^{+0.8}_{-0.8}$   & 12:21:24.50   & +49:18:14.40 & 78   \\
RX J1113-2615	  & 0.730     & $5.6^{+0.9}_{-0.6}$   & 11:13:00.00   & -26:15:49.00 & 103  \\
RX J2302+0844	  & 0.734     & $8.0^{+1.2}_{-1.1}$   & 23:02:48.06   & +08:43:54.72 & 108  \\
MS1137+6625	  & 0.782     & $6.8^{+0.5}_{-0.5}$   & 11:40:22.81   & +66:08:14.50 & 117  \\
RX J1317+2911	  & 0.805     & $4.4^{+1.4}_{-0.9}$   & 13:17:21.84   & +29:11:17.00 & 111  \\
RX J1350+6007	  & 0.810     & $4.5^{+0.7}_{-0.6}$   & 13:50:47.78   & +60:07:13.32 & 58   \\
RX J1716+6708	  & 0.813     & $6.9^{+0.8}_{-0.7}$   & 17:16:48.76   & +67:08:25.81 & 51   \\
MS1054-0321	  & 0.832     & $7.5^{+0.7}_{-0.4}$   & 10:56:58.00   & -03:37:37.30 & 80   \\
RX J1226+3333	  & 0.890     & $12.9^{+1.4}_{-1.3}$  & 12:26:58.20   & +33:32:48.00 & $9.5+31.5$ \\
CL J1415+3612	  & 1.030     & $7.0^{+0.8}_{-0.7}$   & 14:15:11.20   & +36:12:04.00 & 89  \\
RX J0910+5422	  & 1.106     & $6.4^{+1.5}_{-1.2}$   & 09:10:45.41   & +54:22:05.00 & 170  \\
RX J1252-2927	  & 1.235     & $7.3^{+1.3}_{-1.0}$   & 12:52:54.50   & -29:27:18.00 & 188  \\
RX J0849+4452     & 1.261     & $5.3^{+1.7}_{-1.1}$   & 08:48:58.52   & +44:51:55.08 & 185  \\
RX J0848+4453     & 1.273     & $2.4^{+2.5}_{-1.0}$   & 08:48:36.20   & +44:53:47.17 & 185  \\
XMMU J2235-2557$^{\mathrm a}$ & 1.393     & $6.0^{+2.5}_{-1.8}$   & 22:35:20.70   & -25:57:40.70 & 189  \\
\hline
\hline
RCS1419+5326      & 0.620     & $5.0^{+0.4}_{-0.4}$   & 14:19:12.14   & +53:26:11.56 & 56   \\
RCS1107.3-0523    & 0.735     & $4.3^{+0.5}_{-0.6}$   & 11:07:24.08   & -05:23:23.19 & 93   \\
RCS1325+2858      & 0.750     & $1.8^{+1.2}_{-0.6}$   & 13:26:31.04   & +29:03:25.02 & 62   \\
RCS0224-0002      & 0.778     & $5.1^{+1.3}_{-0.8}$   & 02:24:33.61   & -00:02:24.68 & 101  \\
RCS2318.5+0034    & 0.780     & $7.3^{+1.6}_{-1.0}$   & 23:18:30.67   & +00:34:03.03 & 50   \\
RCS1620+2929      & 0.870     & $4.6^{+2.1}_{-1.1}$   & 16:20:10.01   & +29:29:15.41 & 34   \\
RCS2319.9+0038    & 0.900     & $5.3^{+0.7}_{-0.5}$   & 23:19:53.59   & +00:38:09.05 & 62   \\
RCS0439.6-2905    & 0.960     & $1.8^{+0.4}_{-0.3}$   & 04:39:37.76   & -29:04:49.40 & 92   \\
RCS1417+5305      & 0.968     & \ldots                & 14:17:02.13   & +53:05:23.57 & 62   \\
RCS2156.7-0448    & 1.080     & \ldots                & 21:56:41.63   & -04:47:53.47 & 71   \\
RCS2112.3-6326    & 1.099     & \ldots                & 21:12:20.51   & -63:26:13.97 & 68   \\

\hline  

\end{tabular}
\begin{list}{}{}
\item$^{\mathrm[a]}$ The temperature of XMMU J2235-2557 was taken from Mullis et al. \cite{mullis}.
Temperatures of RCS clusters refer to Bignamini et al. submitted.
All other temperatures refer to the values given by Balestra et al. \cite{balestra}.
\end{list}
\end{table*}

      Point sources were removed from all images. The gaps between the detector chips were 
filled assuming Poisson statistics and interpolating with neighboring pixels (dividing images by 
exposure maps is not enough for gap removal because the images are subject to noise propagation).
     The local background was determined in empty regions of the cluster images; when computing the 
background a Gaussian filter with a kernel of 1 was applied to the images to avoid zero pixel
 values. The cluster number counts were determined using the growth curve analysis (B\"ohringer 
et al. \cite{boehringer00}): cluster counts are calculated in circular apertures until the background level 
is reached. The maximum value of the growth curve was taken as the nominal cluster number of counts.
 For each cluster we determined the center-of-mass coordinates, and assigned this 
value to the cluster center.

\subsection{Low-z sample}

     The nearby cluster sample ($0.15<z<0.3$, median $z$=0.21) is presented in Fig.~\ref{Figlowzsample}. The selection criteria for its construction were essentially of observational 
nature since we were interested in low redshift clusters with a size that would properly fit 
within the field of view of Chandra (16 arcmin$^{2}$), and with sufficiently high number of 
counts ($>$ 10 000) to ensure a proper analysis with good enough statistics of the cloned data (see 
section 4.2 for details on the cloning technique). The sample is heterogeneous in the sense that 
we wanted to have clusters with known classification in terms of CC/non-CC; we therefore gathered
 this information from the literature - see Vikhlinin et al. \cite{vikhlinin05}, Bauer et al. \cite{bauer}, Reese et al.
 \cite{reese}, De Filippis et al. \cite{de filippis}, Arabadjis et al. \cite{Arabadjis}, Pratt \& Arnaud \cite{pratt}. According to these
 various sources, clusters A2104, A2218, A2163 and E0657 are obvious non-coolcores whereas A1835
 and ZWCL3146 are rare, high luminosity systems, exhibiting pronounced cool cores. 
   Cluster temperatures obtained with spectroscopic fitting were taken mostly from Baldi et al.
\cite{baldi} and also from Markevitch et al. \cite{markevitch01} \& \cite{markevitch05} for clusters A2163 and E0657, respectively. 
Cores were excised when a cool core was present. All systems have $T > 5$ keV.

\subsection{High-z sample}

     The distant cluster sample (Fig.~\ref{Fighighzsample}) is drawn, 
mostly, from the RDCS (Rosati et al. \cite{rosati98}), with one important addition: a 190 ksec Chandra 
exposure of XMMUJ2235.3-2557 (hereafter XMMUJ2235), with $z$=1.393. This massive object is the 
highest redshift cluster in our sample and the second most distant cluster (Mullis et al. \cite{mullis}) 
confirmed to date (the most distant cluster known so far has redshift 1.45 (Stanford et al. \cite{stanford06}).
The sample ranges from 0.70 to 1.393 in redshift (median $z$=0.83), which spans a lookback 
time between 6.3 and 9.0 Gyrs, for the assumed cosmology. These are all massive clusters, spanning 
approximately a decade in mass, with the exception of RXJ0848+4453 which has a temperature of 
2.4 keV. The remaining clusters in the sample have bulk temperatures which range from 4.4 to 12.9 
keV and were obtained by spectroscopic fitting. Temperature measurements and general Chandra imaging reduction was extensively 
described in Balestra et al. \cite{balestra}. Concerning XMMUJ2235, we quote the temperature obtained by 
Mullis et al. \cite{mullis} using XMM-Newton data, although a more precise spectroscopic analysis of the 
new Chandra data will be available soon (Rosati et al. in preparation).
Cluster number counts range from 250 to 11000 with a median value of 1750 counts.

\section{Surface Brightness profiles}

\subsection{Scaled Surface Brightness profiles}

Galaxy clusters are expected to exhibit self-similarity, which allows us to link their physical 
properties with temperature. This prediction is based solely on gravitational arguments where 
clusters are described as dark matter dominated entities. However, there has been evidence for 
a departure from the self-similar scaling 
scenario (Arnaud et al. 2005b, Chen et al. \cite{chen}): the $L_{X} \propto T^{2}$ scaling relation
expected in the case of gravitational only processes, is at odds with the observed 
$L_{X} \propto T^{2.88}$ relation. This implies the intervention of non gravitational phenomena 
that provide extra energy input (Arnaud \& Evrard 1999). An empirical scaling law was then
 derived which fits to the observations.

   For the comparison of the local and distant clusters, we scaled the surface brightness 
profiles according to, both the empirical and self-similar scaling relations. 
In a self-similar, purely gravitational scenario, the surface brightness scales as 
$S_{X} \propto T^{0.5}$, whereas in an empirical derivation $S_{X} \propto$  $T^{1.38}$ 
(Neumann et al. \cite{neumann}, Arnaud et al. \cite{arnaud02}).

We account for the surface brightness dimming and considered the redshift evolution of 
the $L_{X}-T$ and $R-T$ relations (see e.g. Ettori et al. \cite{ettori}) obtaining: 

\begin{equation}
 S_{X} h(z)^{-3} \propto T^{\alpha} (1+z)^{4} K_{corr} 
\end{equation}
\label{sb}

\noindent where $h(z)$, the cosmological factor, is defined as the Hubble constant normalized to its the 
current value, $H_{z}/H_{0}=\sqrt{\Omega_{m}(1+z)^{3}+\Omega_{\Lambda}}$. The slope $\alpha$ is
 dependent on the used scaling relation and $K_{corr}$ is the K-correction, which is defined in
section 4.2.1.

The way in which scaling relations affect our analysis needs some clarification. 
The purely self-similar gravitational scenario (eg. Navarro, Frenk \& White 1996, 1997), 
which seems to be reasonably well supported by observations, predicts an evolution of the dark
matter halos becoming more compact for fixed enclosed mass with increasing redshift. This 
is reflected in the global appearance of the ICM and the cluster morphology in X-rays, 
including further modification due to changes in the ratio of gas to total mass 
(Arnaud et al. \cite{arnaud02}). This intrinsic evolution of the global cluster appearance is accounted
for by the above mentioned empirical scaling relations. Focusing on the study of cool cores in
clusters we are rather more interested in the local physical conditions at the cluster center
which are for example, characterized by the cooling time of the gas or the total radiative
energy loss of the ICM in the cool core region. This characterization is not subject to the 
above described scaling relations. Thus if we are interested in comparing the rate at which 
the central ICM is radiatively cooling in nearby and distant clusters, we have to compare 
the cluster properties in fixed physical units. We will therefore in the following use the
 empirical scaling relations to reveal changes of the global cluster and cluster ICM
 structure properties, while unscaled relations are used to directly compare physical
conditions in the cooling cores.  

   We adopted a fixed density contrast $\Delta$=500 with respect to the critical density of the
Universe at the cluster redshift ($\rho_{c}(z)$), to study the global properties of clusters. 
The fiducial radius, $R_{500}$, was determined using the $R_{500} - T$ relation in Evrard, 
Metzler \& Navarro 1996, represented by the following fitting function:

\begin{equation}
 R_{500} = h(z)^{-1} 2.48 (kT/10keV)^{0.5} H_{50}/H_{70} 
\end{equation} 
\label{radius}

\noindent Eq. (2) is applicable for clusters with T $>$ 3.5 keV. 

\begin{figure*}
\begin{center}
\includegraphics[width=7.cm, height=6.cm,clip=true]{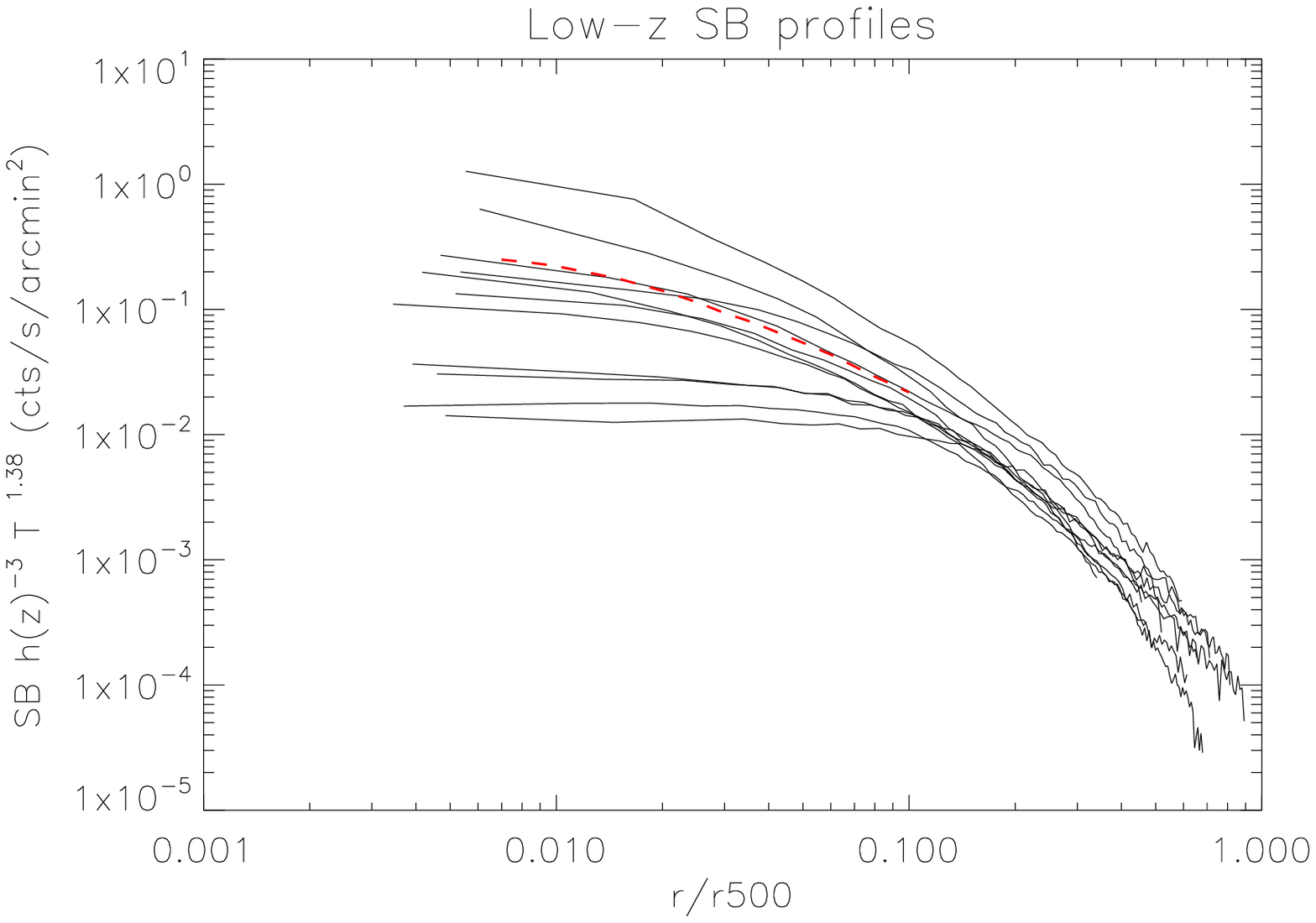}
\includegraphics[width=7.cm, height=6.cm,clip=true]{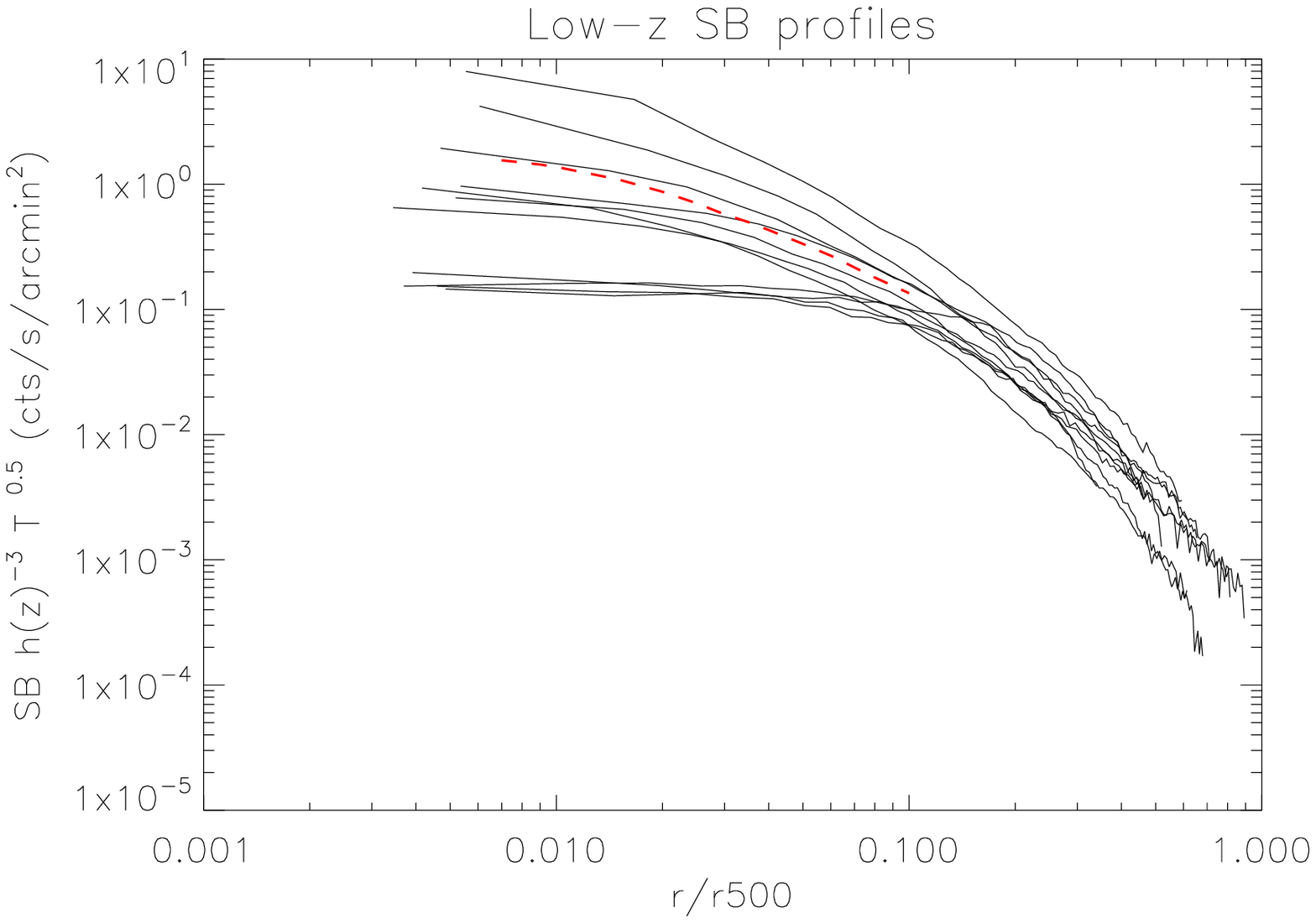}
\includegraphics[width=7.cm, height=6.cm,clip=true]{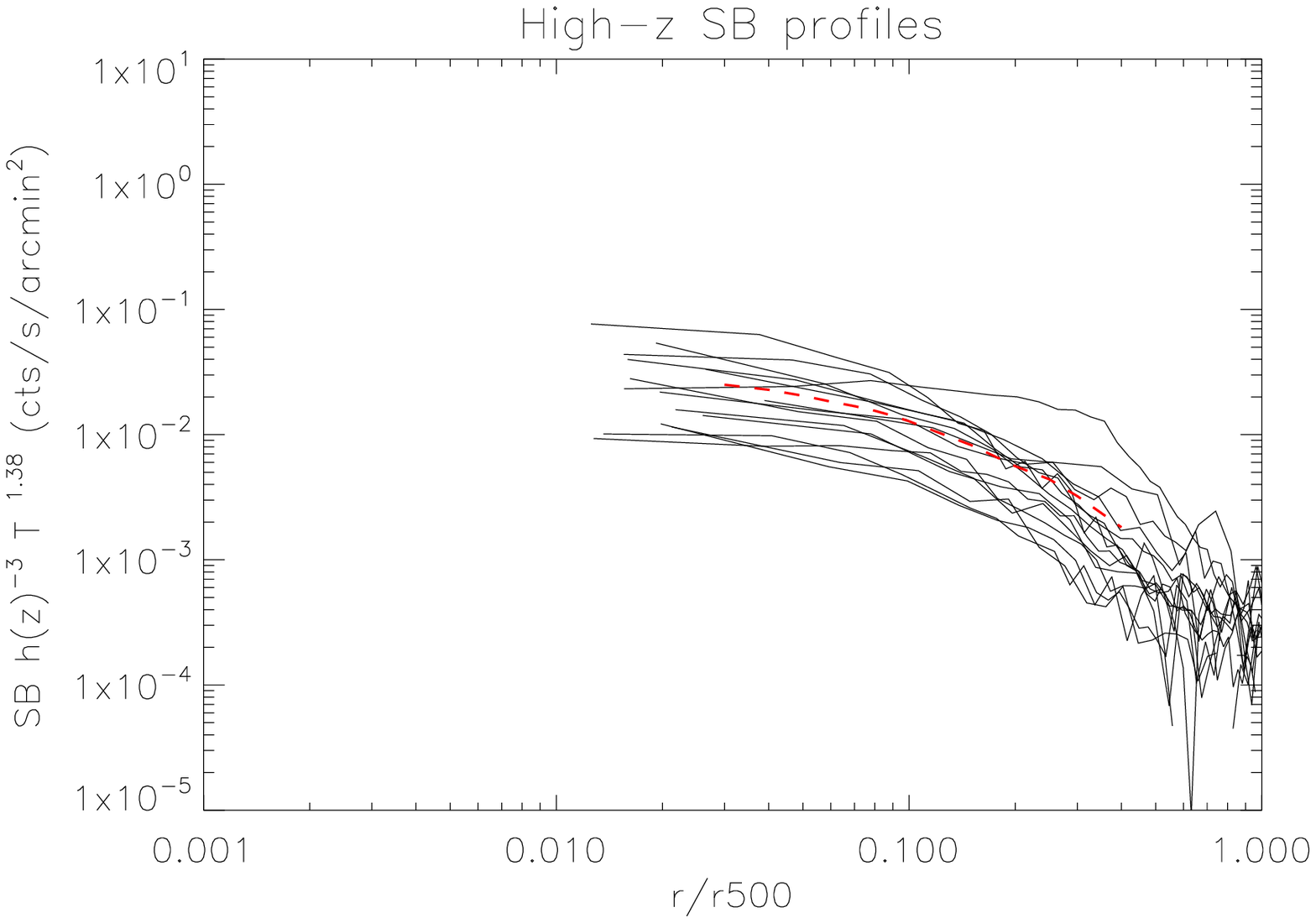}
\includegraphics[width=7.cm, height=6.cm,clip=true]{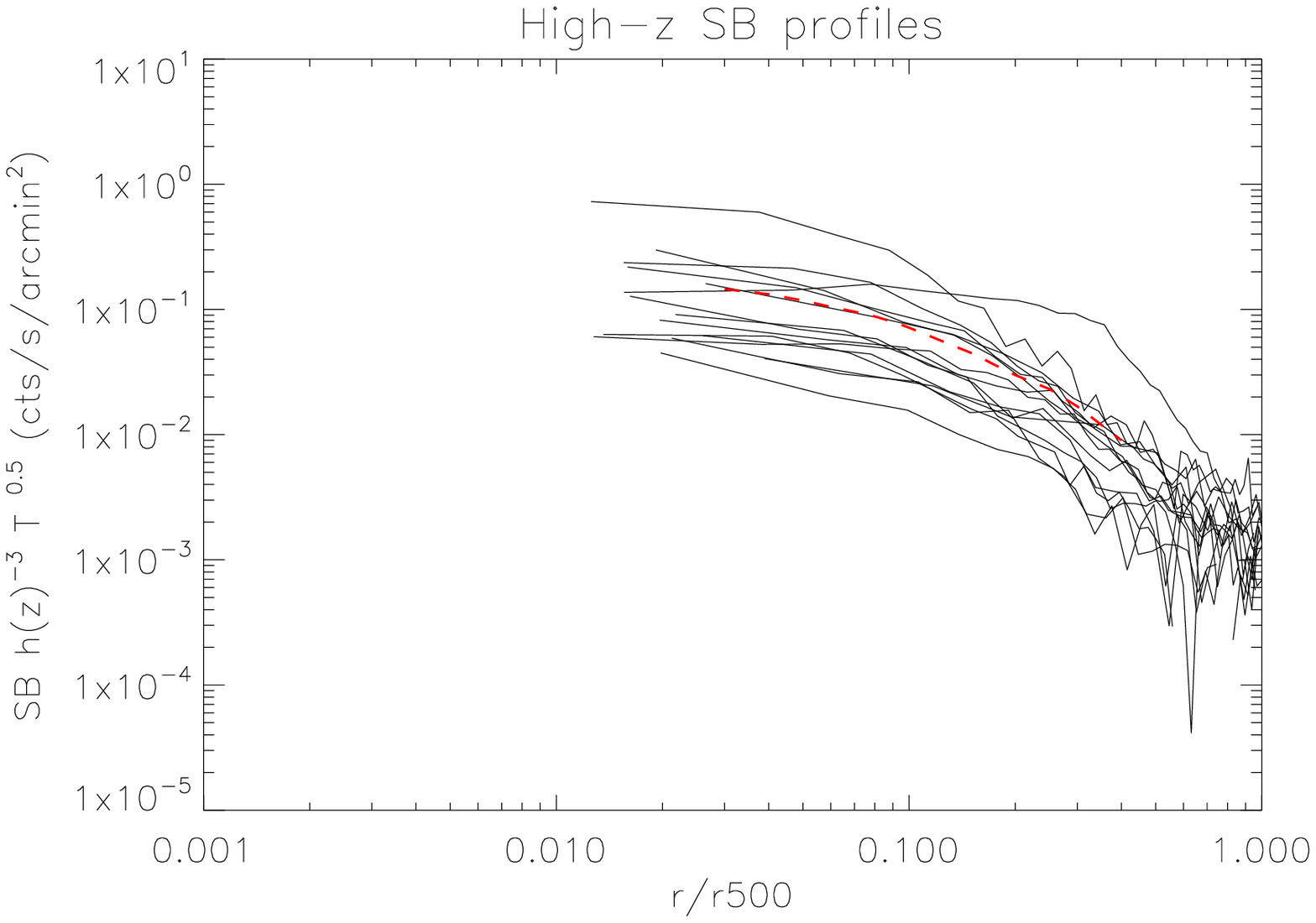}
\caption{$Left$ Empirical scaling of SB profiles - low-$z$ sample ($Top$) and high-$z$ sample 
($Bottom$) $Right$ Self-similar scaling of SB profiles - low-$z$ sample ($Top$) and high-$z$ sample 
($Bottom$). The red dashed lines corresponds to the averaged profiles.}
 \label{Figsbprof}
\end{center}
\end{figure*}

From the top panels in Fig.~\ref{Figsbprof}, it is clear that the low-$z$ profiles converge well for
 $r>0.2.R_{500}$. However, at high-$z$ two outliers are evident in the self-similar scaling 
(bottom-right plot): MS1054-0321 (hereafter MS1054) and RX J1226+3333. The former is a merging
 cluster and the latter is the hottest distant cluster. Interestingly, RX J1226+3333 does not
 appear as an outlier when using the empirical scaling. This just reinforces the better use of
the empirical relation which does not boost the clusters at the extreme (highest temperature)
outside the bulk of the data points in the relation.

A striking difference between the CC and non-CC clusters in the nearby sample is evident in Fig.~\ref{Figsbprof} (top panels). The flatness of the profiles of non-cool core clusters
 A2104, E0657, A2218 and A2163 are in obvious contrast with the remaining profiles which 
have a much higher central surface brightness - up to nearly 2 orders of magnitude larger -
and a steepness which varies according to the strength of the cool core. This result is 
independent of the choice of scaling. At high redshift this marked distinction is not apparent.

The scatter of the surface brightness profiles scaled with $T^{0.5}$ amounts to 50\% 
of the average value in the low-$z$ sample at r=$0.3R_{500}$. In the distant sample the scatter is
boosted by MS1054; if we neglect this outlier we obtain a dispersion around the mean value of 65\% 
at the same radius.

\begin{table*}
\caption{Single and double beta model fitting results: core radius, $r_{c}$, slope $\beta$
 and reduced $\chi^{2}$. Errors are not presented when the parameter value is at the limit
 imposed by the fitting procedure.}             
\label{table:1}      
\centering          
\begin{tabular}{c  l l l | l l l l l}    
\hline
        & \multicolumn{3}{c}{1$\beta$ fit} & \multicolumn{5}{c}{2$\beta$ fit}\\ 
cluster & $\beta$ & $r_{c}$(kpc) & $\chi^{2}_{red}$ & $\beta_{1}$ & $r_{c1}$(kpc) & $\beta_{2}$ & $r_{c2}$(kpc) & $\chi^{2}_{red}$\\ 
\hline                    
A1413	       & 0.473$\pm$0.002 &  56$\pm$1   & 1.821 & 0.755$\pm$0.057  & 70$\pm$5 & 0.900            & 286$\pm$2  & 1.153 \\
A907	       & 0.515$\pm$0.002 &  42$\pm$1   & 6.360 & 0.635$\pm$0.152  & 29$\pm$6 & 0.706$\pm$0.045  & 131$\pm$16 & 1.439 \\
A2104	       & 0.566$\pm$0.014 & 161$\pm$6   & 1.335 & --	          & --	     & 0.542$\pm$0.018  & 162$\pm$7  & 1.447 \\
A2218	       & 0.841$\pm$0.150 & 222$\pm$30  & 1.220 & --               & --	     & 0.606$\pm$0.007  & 159$\pm$4  & 1.326 \\
A2163	       & 1.003$\pm$0.056 & 349$\pm$15  & 1.277 & --	          & --	     & 0.517$\pm$0.003  & 184$\pm$2  & 2.909 \\
A963	       & 0.516$\pm$0.003 &  68$\pm$1   & 1.165 & 0.702$\pm$0.063  & 76$\pm$7 & 0.900	        & 301$\pm$8  & 1.133 \\
A2261	       & 0.540$\pm$0.008 &  66$\pm$3   & 1.147 & 1.200	          & 49$\pm$11& 0.559$\pm$0.006  &  80$\pm$5  & 1.090 \\
A2390	       & 0.474$\pm$0.001 &  46$\pm$1   &17.037 & 0.619 $\pm$0.020 & 49$\pm$2 & 0.900	        & 411$\pm$5  & 2.561 \\
A1835	       & 0.528$\pm$0.001 &  29$\pm$0.3 & 3.530 & 0.699 $\pm$0.050 & 34$\pm$3 & 0.719$\pm$0.016  & 179$\pm$12 & 1.803 \\
ZwCl3146       & 0.573$\pm$0.001 &   2$\pm$1   & 5.012 & 0.639$\pm$ 0.156 & 18$\pm$5 & 0.667$\pm$0.023  &  93$\pm$9  & 1.224 \\
E0657	       & 1.000           & 477$\pm$4   & 1.864 & --	          & --	     & 0.900	        & 433$\pm$3  & 1.919 \\
\hline
RX J1221+4918  & 0.702$\pm$0.067 & 262$\pm$31  & 1.285 & --	          & --	     & 0.763$\pm$0.062  & 281$\pm$28 & 1.321 \\
RX J1113-2615  & 0.668$\pm$0.066 & 107$\pm$17  & 1.178 & 1.000	          & 98$\pm$46& 0.800	        & 188$\pm$84 & 1.048 \\
RX J2302+0844  & 0.451$\pm$0.041 &  73$\pm$16  & 1.039 & 0.470$\pm$0.041  & 68$\pm$8 & --	        &     --     & 1.155 \\
MS1137+6625    & 0.807$\pm$0.041 & 157$\pm$10  & 1.152 & 1.000	          & 77$\pm$25& 0.800	        &  55$\pm$14 & 0.961 \\
RX J1317+2911  & 0.537$\pm$0.094 &  67$\pm$29  & 1.039 & 1.000	          & 52$\pm$38& 0.800	        & 175$\pm$53 & 1.020 \\
RX J1350+6007  & 0.561$\pm$0.095 & 162$\pm$45  & 1.085 & 0.683 $\pm$0.523 & 36$\pm$27& 0.800		& 341$\pm$48 & 0.909 \\ 
RX J1716+6708  & 0.740$\pm$0.062 & 165$\pm$20  & 0.988 & 1.000	          & 28$\pm$6 & 0.695$\pm$0.071  & 154$\pm$25 & 0.930 \\
MS1054-0321    & 1.000           & 443$\pm$5   & 2.826 & --	          & --	     & 0.800		& 360$\pm$5  & 2.924 \\
RX J1226+3333  & 0.710$\pm$0.053 & 131$\pm$16  & 0.851 & 1.000	          & 23$\pm$11& 0.727$\pm$0.063  & 136$\pm$21 & 0.857 \\
CL J1415+3612  & 0.552$\pm$0.026 &  61$\pm$8   & 1.043 & 0.704$\pm$0.094  &140$\pm$40& 0.800		&  35$\pm$12 & 0.954 \\
RX J0910+5422  & 0.997$\pm$0.320 & 177$\pm$57  & 1.017 & 1.000	          & 39$\pm$13& 0.800		& 158$\pm$19 & 1.022 \\
RX J1252-2927  & 0.557$\pm$0.067 &  89$\pm$21  & 1.103 & 1.000	          & 33$\pm$18& 0.601$\pm$0.111  & 108$\pm$41 & 1.114 \\
RX J0849+4452  & 0.537$\pm$0.05	 &  72$\pm$19  & 1.143 & 1.000	          & 29$\pm$14& 0.800		& 127$\pm$17 & 1.043 \\
RX J0848+4453  & 0.590$\pm$0.179 & 122$\pm$69  & 0.823 & 1.000	          & 32$\pm$54& 0.800		& 188$\pm$62 & 1.193 \\
XMMU J2235-2557 & 0.595$\pm$0.028 & 94$\pm$11  & 1.197 & 1.000	          & 33$\pm$18& 0.800		& 155$\pm$11 & 1.090 \\

\hline                  
\end{tabular}
\end{table*}

\subsection{Single and Double Beta model fitting}
  Cluster surface brightness profiles are commonly analyzed with the approximation
of the single isothermal beta model (Cavaliere \& Fusco-Femiano \cite{cavaliere}). We computed 
azimuthally averaged surface brightness profiles in bins of fixed radius, which were
 fitted with a single $\beta$ model:

\begin{equation}
 S(r) = S_{0} (1+(r/r_{c})^2) ^{-3\beta+0.5} + C
\end{equation}
\label{sbeta}

\noindent where $S_{0}$, $r_{c}$, $\beta$ and $C$ are the central surface brightness, 
core radius, slope and constant background, respectively. The fitting procedure was 
performed with a Levenberg-Marquardt least-squares minimization.

   When a cluster harbors a cool core, a single beta model is often inappropriate to 
describe its central excess emission, requiring the use of a $2\beta$ model: 

\begin{equation}
 S(r) = S_{01} (1+(\frac{r}{r_{c1}})^2) ^{-3\beta_{1}+0.5} + S_{02} (1+(\frac{r}{r_{c2}})^2)^{-3\beta_{2}+0.5} + C
\end{equation}
\label{dbeta}

Such a conclusion has been remarked observationally (Jones \& Forman \cite{jones}, Vikhlinin et al. \cite{vikhlinin05}) and 
is also expected from simulations (Burns et al. \cite{burns}). However, this observation cannot be so 
clearly stated for the high-$z$ clusters because the low statistics allows to obtain a good 
$\chi^{2}_{red}$ using either a single or a double $\beta$ function. This has been previously 
noted by Ota \& Mitsuda \cite{ota} who analyzed a sample of 79 clusters with $0.1<z<0.8$, and 
in Ettori et al. \cite{ettori}, where nearly all distant clusters in our sample were analyzed. 
There are however exceptions such as cluster CL J1415+3612 at $z$=1.03, in which the 
single beta model fails to capture the central SB. A double $\beta$ model is qualitatively 
more appropriate to describe the core SB excess, although there are no significant differences 
in the reduced $\chi^{2}$ values of the 1$\beta$ and 2$\beta$ model fit: $\chi^{2}_{red}$=1.043 
in the 1$\beta$ model fit and $\chi^{2}_{red}$=0.954 in the 2$\beta$ model fit. These results make it 
obvious that fitting single/double $\beta$ models to quantify the central surface 
brightness is not conclusive for the high-$z$ sample. 
The $\beta$-model parameters will not be used further in our study to characterize CCs, as we
found them less useful than the following approaches. We nevertheless report our results in Table 3 
for comparison with other work. Since the double-$\beta$ model results give, in general, a 
good representation of the data within the error bars, the model fits also allow for an 
approximate check and reproduction of our analysis.


\section{Surface Brightness Concentration}

\subsection{Introducing $c_{SB}$}

In principle, the most stringent proof for the detection of a cool core is given by 
the temperature decrease in the core with respect to the bulk of the cluster. 
Unfortunately, the poor photon statistics does not permit a spatially resolved 
spectroscopic analysis of the high-$z$ sample.

   Since a central surface brightness excess is a primary indicator (Fabian et al. \cite{fabian84}) 
of the presence of a cool core, we evaluate the core surface brightness in nearby clusters
 by defining a concentration parameter as the ratio of the peak over the ambient SB.
The aim of this approach is to use a single parameter to make a practical 
initial classification of cool cores.
The apertures corresponding to the peak and the bulk were chosen to yield the largest 
separation between the CC and non-CC domains. The $c_{SB}$ parameter was optimized using the 
low-$z$ sample, varying the radius of the central peak and external radius. The optimal $c_{SB}$ 
is found for a peak radius of 40 kpc and a cluster bulk radius of 400 kpc:

\begin{equation}
 c_{SB}= \frac{SB(r<40 kpc)}{SB(r<400 kpc)}
\end{equation}
\label{csb}

   To support this result we carried out extensive Monte Carlo simulations of CC and non-CC 
clusters modelled with, respectively, double and single beta profiles to optimize the choice of 
the internal and external apertures in the definition of $c_{SB}$. We explored the range of 
30-80 kpc around the typical observed size of cool cores in nearby clusters. The chosen 
inner radius of 40 kpc is the one which make the separation between CC and non-CC clusters more obvious. 
The concentration parameter, as defined in eq. (5), was measured in these simulated clusters 
and we find that it efficiently separates the two classes of objects.

 Defining $c_{SB}$ in terms of fractions of a scaled radius such as $R_{500}$
proved not to be suitable because the wide range covered in redshift implies sampling cluster 
regions of various sizes for different redshifts. For example, 0.1$R_{500}$ amounts to 132 
kpc for the most nearby cluster, whereas for the most distant cluster it corresponds
 to 65 kpc. Keeping in mind that the cool core region has been broadly described as the inner 
70 kpc of a cluster, it is evident that using fractions of $R_{500}$ will introduce a 
bias towards lower redshift systems appearing more concentrated. In addition, scaling the radius
 with temperature would only be meaningful if the cluster samples spanned a different temperature 
range and/or if there were poor ($T<3.5$ keV) systems, which is not the case here. Therefore we 
used a physical radius for measuring the concentration index. 

    Having thus defined the empirical $c_{SB}$ parameter using the low-$z$ sample, we 
investigate whether there is a redshift dependence bias associated with this parameter 
which would invalidate the comparison of $c_{SB}$ results between the samples. To this end we developed 
a simulation method which clones the observed low-$z$ clusters to high redshift. Measuring 
the $c_{SB}$ of simulated distant clusters provides a straightforward way to test for a 
redshift dependence. We describe this technique in the following section.

\subsection{Testing $c_{SB}$: simulating distant galaxy clusters with the Cloning method}

   The cloning technique has been applied in an optical study by Bouwens et al. \cite{bouwens}, with 
the purpose of quantifying the evolution of faint galaxies, using low-$z$ galaxies in the Hubble
 Deep Field. By applying a pixel-by-pixel K-correction map to high resolution images of bright
 galaxies, a set of no-evolution deep fields was created, taking into account the space density
 and the cosmological volume.

	We have revised and adapted the cloning method with the purpose of simulating distant 
clusters of galaxies which are taken as standard non-evolving clusters that can be compared 
with distant samples. We clone the low-$z$ sample to the redshifts matching those of the distant
 sample and for consistency in the comparison we normalize the simulated cluster counts to the 
corresponding expected high-$z$ cluster counts, adding as well the observed high-$z$ background. Chandra angular 
resolution is essential for our analysis. Its point spread function (PSF) with a FWHM of $\thicksim$ 1 arcsec 
corresponds to 8 kpc at $z$=1. Considering that the typical physical size for the CC region is
 about an order of magnitude larger than the instrument resolution, there is no need to deconvolve 
the images with the PSF which would make the application of our method very difficult.
One of the main advantages of this technique is that it is model independent, implying that 
the simulations are under control, as no tuning of parameters is required. The assessment of 
the selection effects is rather robust since we only have to rely on the data.

The procedure comprises essentially 2 steps: the flux decrease and the spatial rescaling which the 
template cluster undergoes.

\subsubsection{Flux dimming: luminosity distance and K-correction} 

   The cluster flux decrease is due to both the physical luminosity distance, $D_{L}$, and 
the redshifted energy band. In the adopted cosmology $D_{L}$ is given by 
(see for instance, Carroll, Press \& Turner 1992, Perlmutter et al. \cite{perlmutter}): 
\begin{eqnarray}
 D_{L}=  \frac{c(1 + z)}{H_{0}}\int_{0}^{z} h(z)^{-1} dz 
\end{eqnarray}
\label{da}

The X-ray flux is thus given by  

\begin{equation}
 F_{x}= \frac{L_{x}}{4 \pi D_{L} ^2} \quad
\end{equation}
\label{fx}

\noindent where $L_{X}$ is the X-ray luminosity in the observed frame.
   Therefore, when cloning to a higher redshift, we must compute the 
ratio of the squared luminosity distances.

   As we are using images in a given energy band (0.5-5.0 keV), a bandpass correction - 
K-correction - plays an important role in the flux rescale because we are 
dealing with high redshift values (Hogg et al. \cite{hogg}).
If $f_{\nu}$ is the unabsorbed spectral flux density, we can express the 
K-correction for an energy band [$E_{1},E_{2}$] as:

\begin{equation}
 K_{corr} = \frac{f_{\nu} (z_{1},\frac{E_{1}}{1+z_{1}},\frac{E_{2}}{1+z_{2}})}{ f_{\nu}(z_{2},\frac{E_{1}}{1+z_{1}},\frac{E_{2}}{1+z_{2}}) } \frac{ecf( \frac{E_{1}}{1+z_{1}},\frac{E_{2}}{1+z_{2}},z_{1},N_{H},Z)}{ecf(\frac{E_{1}}{1+z_{1}},\frac{E_{2}}{1+z_{2}},z_{2},N_{H},Z)}
\end{equation}
\label{kcor1}

\noindent where $ecf$ stands for energy conversion factor, a term which relates flux to counts, and depends on 
the cluster redshift, hydrogen column density, $N_{H}$, and metallicity, Z.

\begin{figure}
   \centering
   \includegraphics[width=8cm,height=6.5cm,angle=0,clip=true]{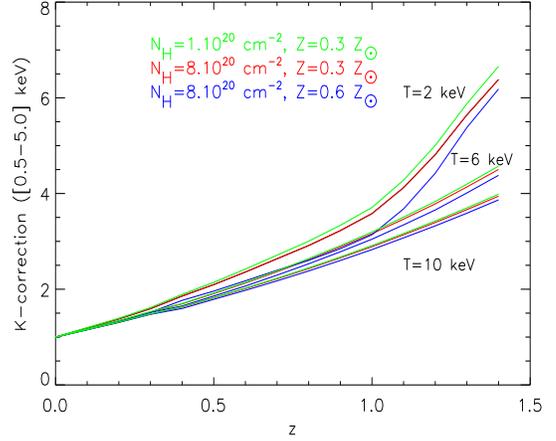}
   \caption{K-correction ratios computed with a MEKAL model: we show the dependence with 
    redshift for temperatures [2,6,10]keV. Red lines refer to an $N_{H}$ of $8.10^{20}$ 
    $cm^{-2}$ and a metallicity of 0.3 solar; blue lines refer to $N_{H}$ = $8.10^{20}$ $cm^{-2}$ 
    and Z/Z$\sun$=0.6 and green lines refer to $N_{H}$ = $1.10^{20}$ $cm^{-2}$ and Z/Z$\sun$=0.3.}   
 \label{Figkcor}
 \end{figure}

 Since the emissivity is not very temperature sensitive in the energy band used for the 
 surface brightness measurements, it is practical and justified to treat the cluster ICM as isothermal.
The K-correction terms are dependent mostly on the cluster redshift (this dependence 
is stronger for $z>0.5$) and temperature. The latter has a steeper effect for poor systems ($T< 3$ keV) 
which is not the case here. 
   Applying single cluster temperatures, K-corrections were computed with XSPEC v11.3.1 (Arnaud \cite{arnaud96}). 
 We used the optically thin MEKAL model spectra (Kaastra et al. \cite{kaastra}) with photoelectric 
absorption (WABS(MEKAL)), setting the metallicity to 0.3 solar and $N_{H}$=$8.10^{20} 
$cm$^{-2}$. The spectra were folded with Chandra responses. The metallicity and the $N_{H}$ have 
a weak influence in the K-correction: at the median redshift 
and temperature of the high-$z$ sample, using a higher value for the metallicity, e.g. $Z/Z\sun$=0.6, 
results in a decrease of the K-correction by $\approx$ 8\%, and a lower $N_{H}$ value of
$10^{20}$cm$^{-2}$ has the marginal effect of raising the K-correction by $\approx$ 2\% - 
see Fig.~\ref{Figkcor} for a description of how the K-correction varies in these parameters.

Eq. (8) thus becomes:

\begin{equation}
 K _{[0.5-5.0]keV}= \frac{\int_{0.5}^{5.0}  f_{\nu}d\nu}{\int_{0.5(1+z)}^{5.0(1+z)} f_{\nu} d\nu}
\end{equation}
\label{kcor2}

\noindent so that $L_{X}=L_{X0}/K_{[0.5-5.0]}$, where $L_{X0}$ is the rest frame 
luminosity of the source.
The calculated correction ratios range from a minimum of 1.5, when we clone a high 
temperature cluster at $z\thicksim0.3$ to a final redshift of 0.7, and may reach a 
maximum of 4 when the cloning redshift is the largest value, i.e, 1.393. 

If a second temperature component related to the cluster core were considered, this would 
increase the K-correction by a factor of 10-20\%.

\subsubsection{Spatial rescaling}

    The spatial rescale which a cluster undergoes when cloned to a higher redshift 
corresponds to the ratio of the angular distances, $D_{A}= D_{L}/(1+z)^2$.
Since the premise of this method is non-evolution of the clusters, it should be 
clear that the size rescale does not account for the shrinking expected when applying
the cosmological factor h(z), which comes about as an intrinsic evolution of clusters.
The spatial resizing was implemented using a gridding linear interpolation.
Note that, as explained above, apart from the redshift resizing, no additional size scaling 
of the kind of eq. (2) was applied, as we compare the cluster cores in physical units.

\begin{figure}[h]
   \centering
   \includegraphics[width=8cm,angle=0,clip=true]{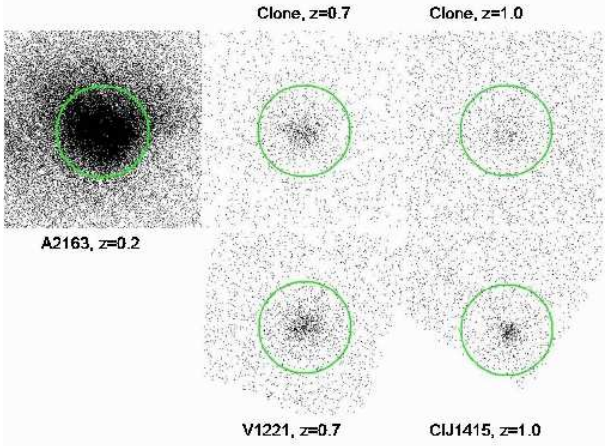}
   \caption{Cloning example: Chandra image of low redshift non-CC cluster, A2163, and 2 
    cloned images (top), compared with a non-cool core cluster at $z$=0.7, RXJ1221 (bottom left),
    and a CC cluster at $z$=1.0, CLJ1415 (bottom right). Green circles have a radius of 2 arcmin.    }
 \label{Figclone}
    \end{figure}
To illustrate this method we show in Fig.~\ref{Figclone} an example of cloning. We clone A2163, a
non-cool core cluster at $z$=0.2, to redshifts 0.7 and 1.0. The observed high-$z$ clusters at the same 
redshifts (RXJ1221 at $z$=0.7 and CLJ1415 at $z$=1.0) are displayed in the lower panels. The simulated X-ray 
images were normalized to have an identical number of counts as the corresponding high-$z$ distant clusters, 
and the same background, which were measured as described in section 2.

A simple consistency check which allows us to evaluate the procedure resides in 
measuring the surface brightness within a physical aperture, both in the real and 
simulated image. For a given energy band we should obtain a ratio of the integrated 
flux which scales as follows:

 \begin{equation}
\frac{S_{X}(z_{i})}{S_{X}(z_{f})} = \frac{(1+z_{f})^{4}}{(1+z_{i})^{4}}  K_{corr}
\end{equation}
\label{check}

We recover perfect agreement of the results from the two approaches (within the expected
numerical errors).

\subsection{The $c_{SB}$ distribution}

\begin{figure} [h]
\begin{center}
\includegraphics[width=6cm,clip=true]{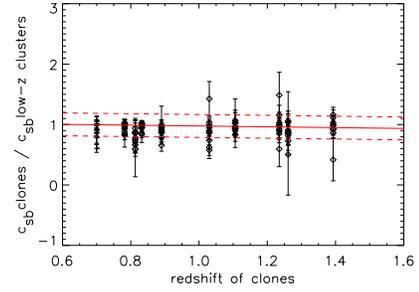}
\caption{Analysis of 10 realizations of cloning the nearby cluster sample to high-$z$: 
we plot the ratio $c_{SB}[clones]/c_{SB}[lowz]$ as a function of cloning redshift. 
Individual error bars refer to 1-sigma confidence level. The red solid line represents the 
linear fit to the data and corresponding 1-sigma errors are shown in dashed line. }
 \label{Figclonesc}
\end{center}
\end{figure}

     We find a very good agreement between the surface brightness concentration of the clones 
and the parent (low-$z$) $c_{SB}$ distribution, implying no redshift dependence in $c_{SB}$. 
This can be exemplified by making 10 realizations of cloning the complete nearby sample to 
$0.7\le z <1.4$ and computing the ratio $c_{SB} [clones] /c_{SB}[low-z] $ as a function of redshift.
 This is illustrated in Fig.~\ref{Figclonesc}: the  ratio of $c_{SB}$ randomly fluctuates around one 
with no trends in redshift, which is confirmed by the weighted mean linear fit to the data.
Moreover, no dependence with temperature was found. This indicates that $c_{SB}$ is an unbiased
 quantity which can be applied to the real high-$z$ sample, therefore we can safely compare the 
nearby with the distant samples using $c_{SB}$.

\begin{figure} [h]
\begin{center}
\includegraphics[width=7.5cm,height=6.5cm,clip=true]{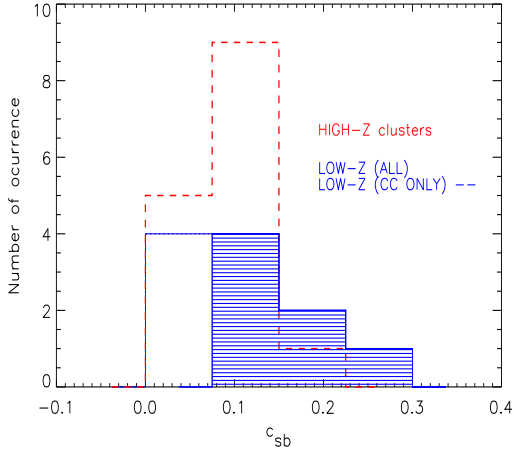}
\caption{Histograms of $c_{SB}$ for the low-$z$ sample (blue solid line with the 
known CC clusters indicated in hatched) and for the high-$z$ sample (red dash line). }
 \label{Figcsb}
\end{center}
\end{figure}

     The distribution of $c_{SB}$ for both the nearby and distant cluster samples in presented in
Fig.~\ref{Figcsb}: the histogram of the low-$z$ sample is shown in blue with the CCC in
blue-hatched and the high-$z$ sample is represented in red. A Kolmogorov-Smirnov (KS) test yields a
probability of 45\% that both $c_{SB}$ distributions were drawn from the same population, 
implying there is no significant difference between the two samples with respect to the 
$c_{SB}$ parameter.
$c_{SB}$ was measured by performing aperture photometry in central regions enclosing 40 and 400 kpc. 
Errors were determined using Poisson statistics - see Table 4 for a listing of these values.

\begin{table}
\caption{Surface brightness concentration, $c_{SB}$ of the nearby and distant
 samples}             
\label{table:4}      
\centering                          
\begin{tabular}{cccc}        
\hline\hline                 
Low-z ID &$c_{SB}$&High-z ID&$c_{SB}$ \\  
\hline                        

A1413	 & $0.095 \pm  0.001$ & RXJ1221+4918 & $ 0.028 \pm 0.004$ \\
A907	 & $0.169 \pm  0.002$ & RXJ1113-2615 & $ 0.095 \pm 0.012$ \\ 
A2104	 & $0.044 \pm  0.001$ & RXJ2302+0844 & $ 0.072 \pm 0.009$ \\
A2218	 & $0.042 \pm  0.001$ & MS1137+6625   & $ 0.096 \pm 0.007$ \\
A2163	 & $0.024 \pm  0.002$ & RXJ1317+2911 & $ 0.123 \pm 0.030$ \\
A963	 & $0.101 \pm  0.002$ & RXJ1350+6007 & $ 0.057 \pm 0.012$ \\
A2261	 & $0.111 \pm  0.003$ & RXJ1716+6708 & $ 0.082 \pm 0.010$ \\
A2390	 & $0.120 \pm  0.001$ & MS1054-0321   & $ 0.016 \pm 0.002$ \\
A1835	 & $0.236 \pm  0.003$ & RXJ1226+3333 & $ 0.083 \pm 0.011$ \\
ZwCl3146 & $0.217 \pm  0.003$ & CLJ1415+3612 & $ 0.151 \pm 0.015$ \\
E0657	 & $0.017 \pm  0.001$ & RXJ0910+5422 & $ 0.101 \pm 0.021$ \\
         &                     & RXJ1252-2927 & $ 0.088 \pm 0.014$ \\
         &                     & RXJ0849+4452 & $ 0.099 \pm 0.023$ \\
         &                     & RXJ0848+4453 & $ 0.064 \pm 0.030$ \\
         &                     & XMMUJ2235-2557 & $ 0.103 \pm 0.012$ \\

\hline                                  
\end{tabular}
\end{table}

\section{Stacked SB profiles in bins of $c_{SB}$}

      Owing to the low number of counts of the distant cluster sample, we resorted to stacking the 
scaled SB profiles which we computed in section 3.1, making use of the $c_{SB}$ results (Table 4). 
From Fig.~\ref{Figcsb} we define three categories of cool cores: non-CC ($c_{SB} < 0.075$), moderate 
($0.075 < c_{SB} < 0.155$) and pronounced ($c_{SB} > 0.155$) CC.

\begin{figure} [h]
\begin{center}
\includegraphics[width=7.5cm,height=6.cm,angle=0,clip=true]{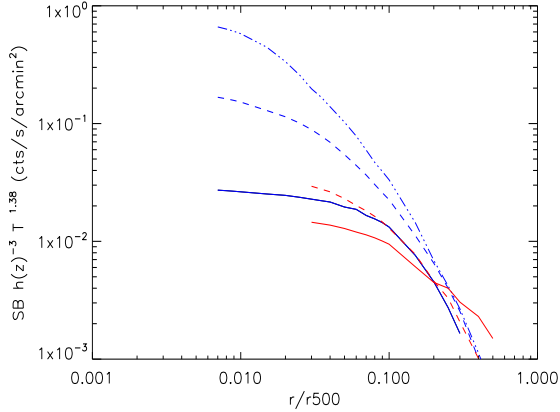}
\end{center}
\vspace*{-0.3cm}
\caption{Stacked surface brightness profiles according to $c_{SB}$ bins: the nearby sample (blue) 
presents 3 bins: non-cool core (solid, bin 1), moderate CC (dash, bin 2) and strong CC (dash-dot, bin 3); 
the distant sample (red) shows 2 bins: non-cool core (solid, bin 1) and moderate CC (dash, bin 2).}
 \label{Figstack}
\end{figure}

 Fig.~\ref{Figstack} shows the stacked profiles of both samples scaled according to the 
empirical scaling law, in units of $R_{500}$. At high-$z$ we are limited by the spatial 
resolution of Chandra up to 0.03 $R_{500}$, and at low-$z$ we are constrained by the background and cluster 
size to probe regions up to $0.5 R_{500}$ (with the exception of A2104 which can be traced only 
as far as 0.3 $R_{500}$). The clusters with complex morphology due to mergers (E0657 at low-$z$ 
and MS1054 at high-$z$) were excluded from the stacking procedure because they introduce strong
 deviations from the average profile, which are related with the difficulty of identifying the
 cluster center. The stacked profiles exhibit different shapes (slope and central emission) 
consistent with Fig.~\ref{Figcsb}, where the low-$z$ sample spans over 3 bins of $c_{SB}$ 
whereas the high-$z$ sample covers the first 2. Both the low and high redshift profiles are in full 
agreement on the outskirts at $r \approx 0.3 R_{500}$ (the deviation of the high-$z$ NCC profile 
(bin1) at large radius is mostly due to noise of the individual profiles), 
but not at the centers where we find a central emission offset: at $r/R_{500}=0.03$ the 
SB of bin1 is 0.023$\pm$0.005 and 0.014$\pm$0.006 cts/s/$arcmin^{2}$ for the low-$z$ 
and high-$z$ samples respectively, with a ratio SB(low-$z$)/SB(high-$z$)=1.6. At the same radius 
the offset is larger for the second bin: SB=0.089$\pm$0.026 cts/s/$arcmin^{2}$ for nearby clusters 
and 0.029$\pm$0.019 cts/s/$arcmin^{2}$ for distant clusters, with a corresponding ratio of 3.0. The 
quoted errors refer to the standard deviations associated with the stacking process, which are larger 
than the combined measurement errors. 

The central surface brightness clearly does not follow the expected predictions from 
scaling relations. As mentioned above, it may be more meaningful to compare the physical state 
of the ICM directly, thus droping the temperature and redshift ($h(z)$) scaling. In this case, 
we obtain a closer match between the low-$z$ and high-$z$ profiles, indicating that the physical 
conditions are similar.

\section{Cooling time analysis}

The radiative cooling of the intracluster gas originates mainly from thermal bremsstrahlung
emission, with additional line and recombination radiation. The cooling time is defined 
as $dt/dlnT_{gas}$ (Sarazin \cite{sarazin}) and allows 
the evaluation of the cooling rate in galaxy clusters. Adopting a gas enthalpy model for the 
cooling function, $t_{cool}$ can be approximated by:

\begin{equation}
t_{cool} = \frac{2.5n_{g}T}{n_{e}^{2} \Lambda(T) }   
\end{equation}
\label{tcool}

\noindent where $\Lambda(T)$, $n_{g}$, $n_{e}$ and T are the cooling function, gas number 
density, electron number density and temperature respectively (Peterson \& Fabian \cite{peterson05}), 
and with $n_{g}$=1.9$n_{e}$. The central cooling time is a sensitive parameter to characterize 
a cool core: when a cool core forms, the central temperature decreases and conversely the 
central density increases, resulting in a small cooling time. Thus we expect that CCC 
present cooling times shorter than the Hubble time or the time since the last major merger event.

\subsection{Density profiles}

\begin{figure} [h]
   \centering
   \includegraphics[width=7.5cm,height=6.5cm,angle=0,clip=true]{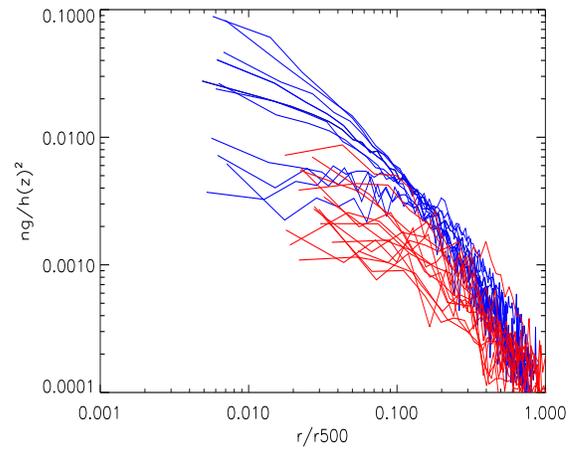}
   \caption{Scaled density profiles: low-$z$ (blue) and high-$z$ (red) clusters}
 \label{Figdensity}
    \end{figure}

Gas density profiles were obtained by deprojection of the surface brightness profiles. 
We corrected the gas density for the cosmological expansion: attending the 
$M_{gas} - T$ and $R-T$ scaling relations (Arnaud et al. 2005a) it follows that $n_{gas}$ 
should be scaled by $h(z)^{-2}$ to account for the more compact shape of distant clusters. 
The density profiles of both cluster populations are presented in Fig.~\ref{Figdensity}. 
Similarly to what we found earlier in the scaled surface brightness profiles, the high-$z$ 
density in the core region is also systematically lower than the central low-$z$ gas density.
This should be due to the h(z) scaling, as explained earlier in section 5.
\vspace*{-0.5cm}

\vspace*{-0.5cm}

\subsection{Cooling time at r=20 kpc}

The cooling time was measured within a central aperture of 20 kpc radius extrapolating 
the density profiles with a $1\beta$ model. We found it preferable to use an aperture 
with fixed physical radius to make a consistent analysis of all clusters, as we are 
limited by the instrument's resolution at high redshift. Using a fraction of a scaled 
radius was not suitable as the redshift range is fairly large, and therefore we would 
probe larger areas at low redshift, introducing a bias. As a result, the strong cool 
cores would have a (artificially) boosted cooling time by a factor $\thicksim$ 2-3.

Since temperature profiles cannot be derived from high-$z$ data, we used the spectral fit 
temperature in the single temperature (MEKAL) model for the cooling time computation. 
As a result, the $t_{cool}$ values in this case are an upper limit because in the 
presence of cool cores the central temperature would be lower, decreasing the cooling time.

\begin{figure} [h]
\begin{center}
\includegraphics[width=7.5cm,height=6.cm,angle=0,clip=true]{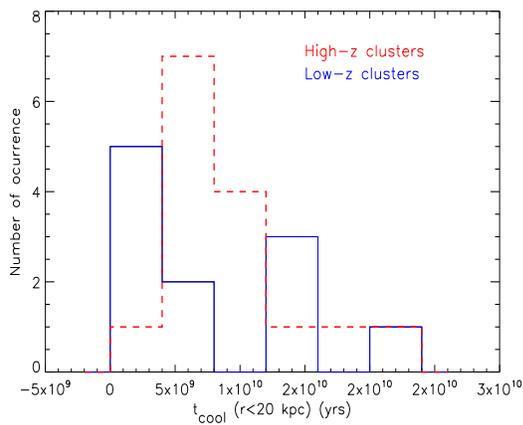}
\end{center}
\vspace*{-0.2cm}
\caption{Distribution of the cooling time, $t_{cool}$, in the low-$z$ (blue) and 
high-$z$ (red) samples. }
\label{Figtcool}
\end{figure}

The cooling time distribution is plotted in Fig.~\ref{Figtcool} and presented in Table 5; 
errors were obtained from the propagation of the errors on the temperature and $ng$. 
Four low-$z$ clusters (A1835, ZwCl3146, A2390 and A907) present a cooling 
time $<$ 3 Gyrs and the 4 nearby non-CC show a $t_{cool} > 10 $ Gyrs. At high-$z$ there is no 
system with $t_{cool} < 3$ Gyrs although the majority (11/15) has $t_{cool} < 10 $ Gyrs, with 
only 4 clusters presenting a $t_{cool} > 10$ Gyrs. 

While similar cooling times or cooling rates in the nearby and distant samples point to similar physical conditions in the central cluster regions, the different ages of the clusters lead to some different 
conclusions of these results. The cooling rates for more than half of the nearby clusters are shorter
than their estimated ages ($\thicksim$10 Gyr) which implies that they approach a steady state of cooling, 
some mass deposition, mass inflow and reheating by central AGN. With the same cooling time distribution, 
only a small fraction of the distant clusters (with ages between 5-8 Gyr) will have a chance to reach
this steady state. If we therefore use the classical definition of cooling flows in which 
$t_{cool} < t_{age}$ also for the distant clusters, we find that the fraction of CCs is 4/15, which is
significantly smaller than the ratio found for nearby clusters (e.g. Peres et al. \cite{peres}, Bauer et al. 
\cite{bauer}, Chen et al. \cite{chen}).

\begin{table}
\caption{Cooling time, $t_{cool}$ measured at 20 kpc, of the nearby and distant
 samples}             
\label{table:5}      
\centering                          
\begin{tabular}{c c c c}        
\hline\hline                 
Low-z ID & $t_{cool}$  (Gyr) & High-z ID & $t_{cool}$  (Gyr) \\  
\hline                        

A1413	 & $5.50^{+0.23}_{-0.17}$  &  RX J1221+4918 &   $ 20.60^{+4.55}_{-4.61}$ \\
A907	 & $2.91  \pm 0.10$        &  RX J1113-2615 &   $ 7.01^{+1.75}_{-1.38}$  \\ 
A2104	 & $15.10 \pm 0.99$        &  RX J2302+0844 &   $ 14.70^{+3.85}_{-3.67}$ \\
A2218	 & $12.10 \pm 1.22$        &  MS1137+6625   &   $ 6.38^{+0.86}_{-0.83}$  \\
A2163	 & $15.10^{+2.06}_{-1.81}$ &  RX J1317+2911 &   $ 8.77^{+4.17}_{-3.36}$  \\
A963	 & $4.52^{+0.32}_{-0.26}$  &  RX J1350+6007 &   $ 10.90^{+3.65}_{-3.24}$ \\
A2261	 & $3.89^{+0.38}_{-0.26}$  &  RX J1716+6708 &   $ 6.39^{+1.34}_{-1.25}$  \\
A2390	 & $2.67 \pm 0.07 $        &  MS1054-0321   &   $ 16.70^{+3.25}_{-2.43}$ \\
A1835	 & $1.13 \pm 0.09$         &  RX J1226+3333 &   $ 4.11^{+0.88}_{-0.83}$  \\
ZwCl3146 & $1.32 \pm 0.08$         &  CL J1415+3612 &   $ 3.90^{+0.72}_{-0.71}$  \\
E0657	 & $20.80 \pm 1.62$        &  RX J0910+5422 &   $ 8.65^{+3.11}_{-2.71}$  \\
         &                         &  RX J1252-2927 &   $ 7.35^{+2.01}_{-1.78}$  \\
         &                         &  RX J0849+4452 &   $ 9.02^{+4.12}_{-3.11}$  \\
         &                         &  RX J0848+4453 &   $ 6.19^{+7.87}_{-4.00}$  \\
         &                         &  XMMU J2235-2557 & $ 4.84^{+2.38}_{-1.82}$ \\

\hline                                  
\end{tabular}
\end{table}

\subsection{The correlation $t_{cool}$ - $c_{SB}$}

\begin{figure}
\begin{center}
\includegraphics[width=8.cm,angle=0,clip=true]{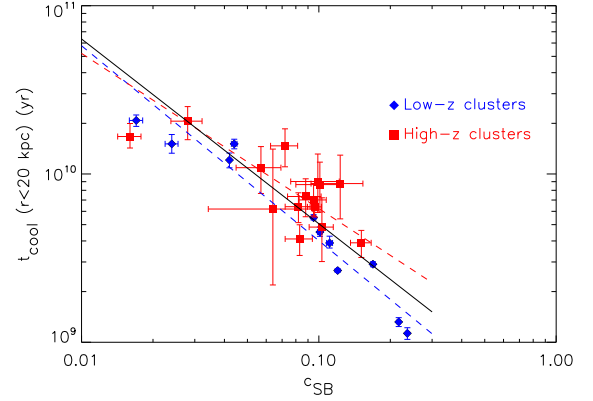}
\end{center}
\vspace*{-0.3cm}
\caption{Correlation between cooling time, $t_{cool}$ and the surface brightness 
concentration, $c_{SB}$, for the low-$z$ (blue diamonds) and high-$z$ (red squares) samples. The black
solid line refers to the composite (both populations) bootstrap fit; the blue and red dashed lines 
refer to separate fits to the nearby and distant samples, respectively. }
 \label{Figcorrelation}
\end{figure}

As we are interested in evaluating the degree of cooling of the central intracluster gas 
with a simple function that can be deduced from imaging data only, we investigated the 
relation between the cooling time and the surface brightness concentration. 
We find a strong negative correlation with a Spearman's $\rho$ rank correlation coefficient 
of -0.84 and a significance of non-correlation $p=8.15$x$10^{-8}$.
Using 1000 bootstrap samples we performed a linear fit in the log-log plane of $t_{cool}$ and $c_{SB}$, 
from which we derived an average slope of the power law function which describes the correlation 
between the two cooling estimators. A composite fit to both cluster samples yields 
$ t_{cool} \propto c_{SB}^{-1.10\pm0.15} $, (Fig.~\ref{Figcorrelation}). Fitting the nearby and 
distant samples separately yields the following correlations: 
$ t_{cool} (high_{z}) \propto c_{SB}^{-0.92\pm0.26} $ and 
$ t_{cool}(low_{z}) \propto c_{SB}^{-1.16\pm0.14}$. 

Using the single fit correlation we obtain the cooling time at the reference 
thresholds $c_{SB}=0.075$ and $c_{SB}=0.155$, as 6.9 Gyrs and 3.1 Gyrs respectively.

\section{X-ray versus Optically selected distant clusters: a comparison using $c_{SB}$ }

In this section we briefly analyze the X-ray surface brightness properties of distant 
galaxy clusters detected in the optical. Specifically, we evaluate the degree of cooling 
using the previously defined $c_{SB}$ parameter and compare with the results obtained for 
the X-ray selected high redshift sample.

The Red Sequence Cluster Survey (RCS) aims at finding and characterizing distant galaxy 
clusters using R and z-band imaging, up to redshift $z$=1.4 (Gladders \& Yee \cite{gladders}).  
    Selecting clusters in the optical allows one to study high redshift poor systems with 
low temperature, which are usually not found in flux limited samples restricted to high 
$L_{X}$ objects at high redshift. 

 Although there are indications of a large scatter in the $L_{X}-T$ relation of the RCS 
sample, such clusters with low temperature (median T = 5 keV) enable us to probe a larger 
dynamical range in the $L_{X}-T$ relation.

    The connection between optical and X-ray characteristics of these distant clusters has
 been under investigation (Hicks et al. \cite{hicks}). Currently, 11 clusters with $0.6<z<1.2$ have 
been observed with Chandra and X-ray properties as temperature and luminosity have been 
derived (Hicks et al. \cite{hicks}, Bignamini et al. submitted). We based our analysis on 
Chandra images reduced by Bignamini et al. - see Table 2 for an overview of 
the sample.

\begin{figure} [h]
\begin{center}
\includegraphics[width=8cm,angle=0,clip=true]{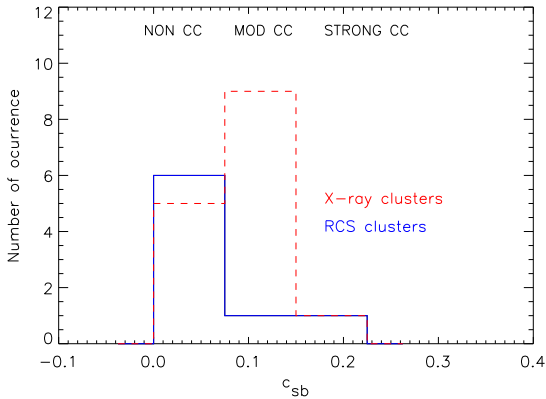}
\end{center}
\vspace*{-0.5cm}
\caption{Histogram of $c_{SB}$ for the RCS sample in solid blue line; the high-$z$ X-ray 
sample is overplotted in red dashed line. }
 \label{Figrcs}
\end{figure}

 We find that 3 of the 11 RCS clusters are barely detected in X-rays and have such a low 
surface brightness that a $c_{SB}$ analysis is not possible. The measurement of the surface
 brightness concentration as described in section 4.3, yields that the 
majority (6 of 8) of the RCS clusters with measurable X-ray emission have $c_{SB} < 0.075$,
 thus falling in the non-CC regime. Of the remaining two: RCS1107 has a $c_{SB} = 0.143$ 
indicating a moderate CC and RCS1419 has $c_{SB} = 0.185$, suggesting a strong CC - see 
histogram in Fig.~\ref{Figrcs}. The $c_{SB}$ distribution of the X-ray distant clusters 
is overplotted in the same figure (red line) for direct comparison. The Kolmogorov-Smirnov
 statistic between the two $c_{SB}$ distributions is 0.4, with a probability of the samples
 being drawn from the same distribution of 20\%. With such low statistics, the KS test does
 not provide a sensitive distinction on these two datasets. 

The outcome of this analysis based on our small samples suggests that high-$z$ optically selected 
clusters may have a lower fraction of cool cores with respect to clusters detected in X-rays.
Larger samples are needed to draw conclusions on possible different physical conditions of the ICM 
in optically selected clusters, which may have not reached a final state of virialization

\vspace*{-0.2cm}

\section{Discussion and Conclusions}

 In this paper we investigate the detection and quantify the strength of cool cores in the 
high redshift cluster population.

 We obtained azimuthally averaged SB profiles and scaled them, in this manner accounting for 
 different cluster temperatures. The segregation of cool cores at low redshift using this method 
 is evident in Fig.~\ref{Figsbprof} (top panels). However, this segregation is not obvious in the
distant sample, and the use of single/double $\beta$ model fitting does not provide conclusive results.

 A surface brightness concentration index, $c_{SB}$, was then defined and measured in the local 
sample, which efficiently separates the non-cool core from the cool core regime. 
This is in agreement with previously published results on the cuspiness of these clusters. 
In Hashimoto et al. \cite{hashimoto}, a similar surface brightness concentration was defined, using 
the ratio of elliptical aperture radii enclosing, respectively, 30 and 100 \% of the cluster
 surface brightness. This parameter was 
applied to a sample of 101 clusters with $0.05<z<1$, these including 4 of the 11 nearby 
clusters and 12 of the 15 distant clusters we study here. Comparing the results from both
 measurements for individual clusters we find a good agreement with our results, with the 
 exception of A1835. 

 We also developed a robust parameter free method to simulate galaxy clusters 
at redshift $z\ge0.7$, making use of the heterogeneous nearby ($0.15<z<0.3$) sample. These 
simulated clusters serve as a benchmark for non evolving clusters, which allow us to test 
the suitability of indicators of cool cores at high redshift. Using our cloning technique, 
we verified the redshift independence of $c_{SB}$, thus making
 it a simple, unbiased quantity to study cool cores in the distant cluster sample 
(Fig.~\ref{Figcsb}). The analysis of the concentration and radial profiles allows us to 
define three $c_{SB}$ bins which distinguish different regimes: non-CC ($c_{SB}<0.75$), 
moderate CC ($0.75<c_{SB}<1.55$) and strong CC ($c_{SB}>1.55$). By stacking the profiles in 
these bins (Fig.~\ref{Figstack}), we obtained a robust classification of cool cores which 
indicate the range of different cooling rates which the two cluster samples span. This is 
particularly useful for the distant sample which is affected by low X-ray count statistics.

From the overall surface brightness analysis we conclude that the majority (10/15) of the 
high-$z$ clusters present mild cool cores, similar to those found in the nearby sample. 

    In addition to the surface brightness analysis, we measured the clusters' cooling time 
in a central region with 20 kpc radius, using a single global temperature. Analysing the 
reference, low-$z$ sample, we conclude that strong cool cores 
present $t_{cool} < 3 $ Gyrs, whereas moderate CC have $ 3 < t_{cool} < 10 $ Gyrs and 
finally non-CC show $t_{cool} > 10 $ Gyrs. The majority (11/15) of the high-$z$ clusters 
are characterized by $ 3 < t_{cool} < 10 $ Gyrs with only 4 systems showing $t_{cool} > 10 $ 
Gyrs. We found no distant clusters with $t_{cool} < 3 $ Gyrs. Similar cooling time bins were 
provided by Bauer et al. \cite{bauer}, to discriminate between strong, mild and non-cool core systems. 
These authors measured the central cooling time and also the cooling time at 50 kpc: good 
agreement on their $t_{cool}$ at 50 kpc and our $t_{cool}$ at 20 kpc is found for the common 
objects in the intermediate-z sample of Bauer et al. \cite{bauer} (4/6 objects within $1\sigma$ errors).
Even so the distribution of the cooling time is quite similar in the nearby and distant 
samples, a smaller fraction of distant clusters has $t_{cool} < t_{age} \thicksim t_{Hubble}$,
and therefore the fraction of distant CCs following the classical cooling flow distribution 
is smaller than for local clusters.

   We find a strong correlation between $t_{cool}$ and $c_{SB}$, which allow us to reliably 
relate a physical quantity with a phenomenological parameter. This correlation is described 
by a power law fit:  $ t_{cool} \propto c_{SB}^{-1.10\pm 0.15} $ (Fig.~\ref{Figcorrelation}). 
The correlation does not change significantly if we fit the two samples separately. The low-$z$ 
clusters show a slightly steeper slope ($-1.16 \pm 0.14$) when compared with the high-$z$ slope 
($-0.92 \pm 0.26$).

    Investigating galaxy cluster properties at high redshift is difficult as these systems 
have small angular size, requiring high resolution instruments and long integration time, 
particularly for spectroscopy. It is therefore understandable that few attempts have been made 
to probe the centers of distant galaxy clusters. Vikhlinin et al. \cite{vikhlinin06} presented a study on the 
evolution of CCC in the redshift range [0.5-0.9] which is compared to a local sample. They find 
evidence for evolution of the CC fraction, with a lack of cool cores in the distant sample. 
They use a different indicator for the surface brightness cuspiness which might have a lower 
sensitivity to moderate cool core systems. A meaningful comparison with our study would require 
a detailed analysis of the different cool core estimators.

 One often finds that massive, luminous galaxies containing AGN, lie at the centers of 
cool core clusters (Eilek \cite{eilek}). In an attempt to establish such a correlation we searched 
for cD galaxies in the cluster samples, examining available optical images. We find that, 
in the low-$z$ sample, all CCC indeed host a bright, massive galaxy, but also A2104, a 
non-cool core cluster, possesses a cD galaxy with an associated AGN (Liang et al. \cite{liang}, 
Martini et al. \cite{martini}). Could this be a sign of an AGN heating overshoot which could have 
contributed to destroy the cool core? At high redshift there is no obvious trend: the 
majority of the clusters possess a cD galaxy but it does not correlate with $c_{SB}$. 

 In the established scenario of hierarchical structure formation, galaxy clusters develop
 through mergers and by accretion of neigbouhring smaller objects.
Numerical simulations (Cohn \cite{cohn}) predict higher cluster merger rates with increasing 
redshift, where the fraction of clusters with recent mass accretion due to mergers can 
be doubled at $z$=0.7, with respect to the local abundance. This framework provides a 
possible mechanism for preventing the formation of prominent cool cores at high redshift. 
However, since high-$z$ clusters are younger, other mechanisms may cause a delay in the 
formation of cool cores due to some internal energy release, such as AGN activity and 
star formation processes. The observed absence of pronounced CC at $z\ge0.7$ is therefore 
plausible in the current cosmological framework. 

We also investigated possible sample selection effects in the search for high redshift 
cool cores by studying 11 optically selected clusters from the RCS sample, which have a 
median temperature of 5 keV. We found that, unlike the X-ray sample, the majority of 
clusters in the RCS subsample have $c_{SB}<0.075$, i.e. they lie in the non-cool core 
category.

    To further extend this analysis we need to construct a complete low redshift sample, 
possibly with the same selection function as the distant sample, to be able to trace the 
evolution of the abundance of cool cores. A good prospect for this study is the local 
representative cluster sample XMM-Large Programme (B\"ohringer et al. \cite{boehringer07}), which 
comprises 33 clusters in the redshift range $z$=0.055 to 0.183, drawn from the REFLEX 
sample (B\"ohringer et al. \cite{boehringer01}). An intermediate redshift sample (z=0.3-0.6) from a 
similar project is also available (P.I. M. Arnaud) which provides an intermediate step 
in the assessment of the cool core evolution with redshift.

Clearly, having a high-$z$ data set covering a larger volume would also be very advantageous.
Serendipitous cluster surveys currently underway, such as the XCS (Romer et al. \cite{romer}), 
the XDCP (Mullis et al. \cite{mullis}, B\"ohringer et al. \cite{boehringer05}) and XMM-LSS (Pierre et al. \cite{pierre}), 
will yield suitable distant samples for these studies in the near future.

A comparison of our results with cosmological simulations would also provide additional 
new constraints to cluster formation scenarios. In fact, new insights on the evolution of 
simulated CCs and NCCs have recently been published in Burns et al. \cite{burns}, which are in good agreement 
with our findings.

\begin{acknowledgements}
      We would like to thank A. Baldi for giving access to the low redshift Chandra data.
 JSS thanks Y-Y. Zhang, G. Pratt, D. Pierini and I. Balestra for useful discussions. JSS 
 is supported by the Deutsche Forschungsgemeinschaft under contract BO702/16-2. PT and SE 
acknowledge the financial contribution from contract ASI-INAF I/088/06/0.
\end{acknowledgements}

\end{document}